\documentclass[a4,12pt]{article}
\usepackage{graphics}
\usepackage[usenames]{color}
\usepackage[dvips]{epsfig}
\usepackage{mathrsfs}
\usepackage{amsmath,amssymb}
\newcommand{\be}{\begin{equation}}
\newcommand{\ee}{\end{equation}}
\newcommand{\bea}{\begin{eqnarray}}
\newcommand{\eea}{\end{eqnarray}}
\newcommand{\bean}{\begin{eqnarray*}}
\newcommand{\eean}{\end{eqnarray*}}
\newcommand{\ba}{\begin{array}}
\newcommand{\ea}{\end{array}}

\newcommand{\bc}{\begin{center}}
\newcommand{\ec}{\end{center}}
\newcommand{\bi}{\begin{itemize}}
\newcommand{\ei}{\end{itemize}}

\newcommand{\norsl}{\normalsize\sl}
\newcommand{\norsc}{\normalsize\sc}

\newcommand{\ra}{\rightarrow}
\newcommand{\lra}{\longrightarrow}

\newcommand{\lsim}{\raisebox{-0.07cm   }
{$\, \stackrel{<}{{\scriptstyle\sim}}\, $}}
\newcommand{\grtsim}{\raisebox{-0.07cm   }
{$\, \stackrel{>}{{\scriptstyle\sim}}\, $}}

\newcommand{\aqt}{{\mathscr{A}_{TT}(Q_T)}}
\newcommand{\aqtn}{\mathscr{A}_{TT}}

\begin{document}

\title{Soft gluon corrections to double transverse-spin asymmetries 
for small-$Q_T$ dilepton production\\
at RHIC and J-PARC}   

\author{
{\norsc  Hiroyuki Kawamura${}^a$, Jiro Kodaira${}^b\footnote{Deceased.}
$ and Kazuhiro Tanaka${}^c$}\\
\\
\norsl  ${}^a$ Radiation Laboratory, RIKEN, Wako 351-0198, 
Japan \\
\norsl  ${}^b$ Theory Division, KEK, Tsukuba 305-0801, Japan
\\
\norsl  ${}^c$ Department of Physics, Juntendo University,
    Inba, Chiba 270-1695, Japan}
\date{}
\maketitle

\begin{abstract}
We calculate the double transverse-spin asymmetries, $\aqt$,  
in transversely polarized Drell-Yan process 
at the transverse-momentum $Q_T$ of the produced lepton pair.
We perform all-order resummation of the logarithmically enhanced contributions 
in the relevant Drell-Yan cross sections at small $Q_T$,
which are due to multiple soft gluon emission in QCD,
up to the next-to-leading logarithmic accuracy.   
The asymmetries $\aqt$ to be observed in polarized experiments at RHIC and J-PARC 
are studied numerically as a function of $Q_T$.
We show that the effects of the soft gluon resummation to the polarized 
and unpolarized cross sections largely cancel in $\aqt$, 
but the significant corrections still remain and are crucial for making a reliable
QCD prediction of $\aqt$. In particular, the soft gluon corrections enhance
$\aqt$ considerably in the small $Q_T$ region compared with 
the asymmetry in the fixed-order $\alpha_s$ perturbation theory.
We also derive a novel asymptotic formula which embodies those
remarkable features of $\aqt$ at small $Q_T$ in a compact analytic form
and is useful to extract the transversity $\delta q(x)$ from the experimental data.
\end{abstract}

\newpage
\section{Introduction}

In recent years, a variety of experiments has been devoted to explore 
the spin-dependent phenomena in hard processes.
Especially, experiments 
with transvesely polarized hadrons have opened a new window to study 
rich structure of perturbative/nonperturbative dynamics of QCD 
associated with the transverse spin \cite{BDR:02}. 
One of the fundamental quantities which newly enter into play 
is the chiral-odd, twist-2 parton distribution, called the transversity  
$\delta q(x)$; it represents the 
distribution of transversly
polarized quark inside transversly polarized nucleon, i.e, 
the partonic structure of the nucleon which is complementary to that
associated with the other twist-2 distributions, such as the familiar 
density and helicity distributions $q(x)$ and $\Delta q(x)$.
However, $\delta q(x)$ has not been well-known so far.
This is because $\delta q(x)$ cannot be measured in inclusive DIS 
in contrast to $q(x)$ and $\Delta q(x)$; the chiral-odd nature 
requires a chirality flip, so that 
$\delta q(x)$ must be always accompanied with another chiral-odd
function in physical observables.
It is very recent that the first global fit of 
$\delta q(x)$ is given \cite{Anselmino:07}
using the semi-inclusive DIS data, in combination with 
the $e^{+}e^{-}$ data for the associated chiral-odd (Collins) fragmentation function.

Transversely polarized Drell-Yan (tDY) process, 
$p^{\uparrow}p^{\uparrow}\lra l^+l^-X$,
is another promising process to
access the transversity $\delta q(x)$. 
Based on QCD factorization, the spin-dependent cross section
$\Delta_T d \sigma \equiv (d\sigma^{\uparrow\uparrow}-d\sigma^{\uparrow\downarrow})/2$
is given as a convolution,
$\Delta_T d\sigma = \int d x_1 d x_2 \delta H (x_1, x_2 ; \mu_F^2)$
$\Delta_T$$d \hat{\sigma} (x_1^0 /x_1, x_2^0 /x_2 ; Q^2, \mu_F^2/Q^2)$,
where $Q$ is the dilepton mass, $\mu_F$ is the factorization scale, 
\begin{equation}
\delta H(x_1,x_2;\mu_F^2)=\sum_{q} e_q^2 
\left[
\delta q(x_1 ,\mu_F^2)\delta\bar{q}(x_2,\mu_F^2)
+\delta \bar{q}(x_1 ,\mu_F^2)\delta q(x_2, \mu_F^2)
\right],
\label{tPDF}
\end{equation} 
is the product of transversity distributions of the two nucleons, 
summed over the massless quark flavors $q$
with their charge squared $e_q^2$,
and $\Delta_T d \hat{\sigma}=(d\hat{\sigma}^{\uparrow\uparrow}
-d\hat{\sigma}^{\uparrow\downarrow})/2$ 
is the corresponding partonic 
cross section. 
$x_1^0 = \sqrt{\tau}\ e^y , x_2^0 =\sqrt{\tau}\ e^{-y}$ 
are the relevant scaling variables, where $\tau =Q^2/S$, and
$\sqrt{S}$ and $y$ are the total energy and dilepton's rapidity in the 
nucleon-nucleon CM system.
At the leading twist level, the gluon does not contribute 
to the transversely polarized, chiral-odd process, corresponding 
to helicity-flip by one unit.
The unpolarized cross section,
$d \sigma \equiv (d\sigma^{\uparrow\uparrow}+d\sigma^{\uparrow\downarrow})/2$, 
obeys factorization similar as $\Delta_T d \sigma$, 
in terms of $H(x_1, x_2 ; \mu_F^2)$ that is given
by (\ref{tPDF}) with $\delta q \rightarrow q$ and $\delta \bar{q} \rightarrow \bar{q}$,
and additional functions involving the gluon distribution that   
comes in as higer-order $\alpha_s$ corrections. 
Therefore, the double-spin asymmetry in tDY, $A_{TT}\equiv \Delta_T d \sigma/ d \sigma$,
in principle provides clean information on the transversity $\delta q(x)$. 
At the leading order (LO)
in QCD perturbation theory, $x_{1,2}^0$ coincide with the momentum fractions 
carried by the incident partons, e.g., 
$\Delta_T d \hat{\sigma}\propto \delta(x_1 -x_1^0)\delta(x_2 -x_2^0)$,
so that 
\cite{Ralston:1979ys,BDR:02}
\begin{equation}
A_{TT}=\frac{\Delta_T d \sigma}{d \sigma}
=\frac{1}{2}\cos(2 \phi) \frac{\delta H(x_1^0, x_2^0; Q^2)+\cdots}
{H(x_1^0, x_2^0; Q^2)+\cdots} \ ,
\label{eq:att}
\end{equation}
where $\phi$ denotes the azimuthal angle of one of 
the leptons with respect to the incoming nucleon's spin axis, and the ellipses
stand for the QCD corrections of NLO or higher.
The $\cos(2\phi)$ dependence
is characteristic of the spin-dependent cross section $\Delta_T d \sigma$
of tDY \cite{Ralston:1979ys}.

$A_{TT}$ to be observed in tDY at RHIC-Spin experiment 
was calculated by Martin et al.~\cite{MSSV:98} including the NLO QCD corrections.
The results are somewhat discouraging in that the corresponding $A_{TT}$ are 
at most a few percent \cite{MSSV:98}.~\footnote{In \cite{MSSV:98},
the corresponding asymmetries are defined through certain integration over $\phi$,
and equal (\ref{eq:att}) with the formal replacement $\cos(2\phi) \rightarrow 2/\pi$.}
The reason is twofold (see (\ref{eq:att})):
(i) tDY in $pp$ collisions probes the product of the quark transversity-distribution
and the antiquark one as (\ref{tPDF}),
and the latter is likely to be small;
(ii) the rapid growth of the unpolarized sea-quark distributions in 
$H(x_1^0, x_2^0; Q^2)$ 
is caused by the DGLAP evolution in the low-$x$ region that is typically probed at RHIC,
$\sqrt{S}=200$ GeV, $Q \lesssim 10$ GeV, and $\sqrt{\tau}\lesssim 0.05$.
Thus, small $A_{TT}$ at RHIC appears to be 
rather general conclusion (see also \cite{WV:98}).

We note that those previous NLO studies of $A_{TT}$ of (\ref{eq:att})
correspond to tDY with the transverse-momentum $Q_T$ of the produced
lepton pair unobserved, and use the cross sections $\Delta_T d \sigma,
d\sigma$ integrated over $Q_T$ in (\ref{eq:att}).
However, in view of the fact that most of the lepton pairs are actually 
produced at small $Q_T$ in experiment,
it is important to examine the double transverse-spin asymmetries at a measured $Q_T$,
in particular its behavior for small $Q_T$.
This is defined similarly as (\ref{eq:att}) using the
``$Q_T$-differential'' cross sections, and we denote it as $\aqt$
distinguishing from the conventional $Q_T$-independent $A_{TT}$.
In fact, participation of the new scale $Q_T (\ll Q)$
causes profound modifications of the relevant theoretical framework.
For example, now the numerator and the denominator 
of $\aqt$ may involve the parton distributions associated with the scales $\sim Q_T$,
such as $\delta H(x_1^0 , x_2^0; Q_T^2)$ and $H(x_1^0 , x_2^0; Q_T^2)$, respectively.
For $H(x_1^0 , x_2^0; Q_T^2)$, the low-$x$ rise of the unpolarized 
sea-quark distributions, mentioned in (ii) above,
is milder compared with $H(x_1^0 , x_2^0; Q^2)$.
Thus, if the former components play dominant roles compared with the latter 
in the denominator of $\aqt$ by certain partonic mechanism, 
$\aqt$ for small $Q_T$ region at RHIC can be larger than $A_{TT}$;
the necessary partonic mechanism is indeed
provided as the large logarithmic contributions
of the type $\ln (Q^2 /Q_T^2 )$, which is
another remarkable consequence of the new scale $Q_T$:
the small transverse-momentum $Q_T$ of the final lepton pair is provided by
the recoil from the emission of soft gluons 
which produces the large terms behaving 
as $\alpha_s^n\ln^m(Q^2/Q_T^2)/Q_T^2 ~(m=0, 1, \ldots, 2n-1)$  
at each order of perturbation theory for the tDY cross sections.
Actually, such enhanced ``recoil logarithms'' spoil
the fixed-order perturbation theory, and  
have to be resummed to all orders in $\alpha_s$ to make a reliable 
prediction of the cross sections at small $Q_T$.
Recently, we have worked out the corresponding ``$Q_T$-resummation'' 
for the tDY cross sections
up to next-to-leading logarithmic (NLL) 
accuracy, which corresponds to summing up exactly the first
three towers of logarithms, $\alpha_s^n\ln^m(Q^2/Q_T^2)/Q_T^2$ with $m=2n-1,2n-2$
and $2n-3$, for all $n$ \cite{KKST:06}.
Utilizing this result, in the present paper,
we develop QCD prediction for $\aqt$ as a function of $Q_T$.

We will demonstrate that the soft gluon corrections are significant
so that 
$\aqt$ in the small $Q_T$ region is considerably 
large compared with the known value for $A_{TT}$.~\footnote{For the
impact of the $Q_T$ resummation on the spin asymmetries in
semi-inclusive deep inelastic scattering, see \cite{KNV:06}.}
In addition to $\aqt$ in tDY at RHIC,
we calculate $\aqt$ to be observed at J-PARC when the polarized beam is realized
\cite{Dutta}. 
The latter case is also interesting because 
the fixed target experiments at J-PARC 
probe the parton distributions in the medium $x$ region ($\sqrt{S}=10$
GeV, $Q \grtsim 2$ GeV, and $\sqrt{\tau}\grtsim 0.2$), and thus 
large asymmetries are expected even for the $Q_{T}$-independent $A_{TT}$ \cite{CDL:06} 
(see (ii) above). 
We also find that $\aqt$ for $Q_T \approx 0$ deserves special attention from theoretical
as well as experimental point of view, 
and derive a compact analytic formula for $\aqtn(Q_T \approx 0)$. 

The paper is organized as follows. In Sec.~2, 
the $Q_T$-resummation formula for the tDY cross sections is introduced,
and all ingredients necessary for
calculating the $Q_T$-dependent asymmetries $\aqt$ including the NLL 
resummation contributions are explained.
In Sec.~3, numerical results of $\aqt$ at RHIC and J-PARC
are presented. Sec.~4 is devoted to the discussion of
analytic formula of 
$\aqt$ at $Q_T \approx 0$
using the saddle-point method. 
Conclusions are given in Sec.~5.

\section{Resummed cross section and asymmetry for tDY}

Throughout the paper we employ the $\overline{\rm MS}$ factorization and 
renormalization scheme with the corresponding scales, $\mu_F$ and $\mu_R$.
We first recall basic points of the fixed-order calculation of the spin-dependent,
$Q_T$-differential cross sections of tDY \cite{KKST:06}.
In the lowest-order approximation via the Drell-Yan mechanism, 
the lepton pair is produced with vanishing $Q_T$,
so that the corresponding partonic 
cross section is proportional to $\delta (Q_T^2)$. 
The one-loop corrections to the partonic cross section 
involve the virtual gluon corrections, and the real gluon emission contributions, 
$q + \bar{q} 
              \to l + \bar{l} + g$;
in the latter case, the finite $Q_T$ of the lepton pair is provided by the recoil 
from the gluon radiation.
Those have been calculated in the dimensional regularization \cite{KKST:06}, and 
the differential cross section of tDY is obtained as
\begin{eqnarray}
\frac{\Delta_T d \sigma^{\rm FO}}{d Q^2 d Q_T^2 d y d \phi}
= 
\cos{(2 \phi )}
\frac{\alpha^2}{3\, N_c\, S\, Q^2}
 \left[ \Delta_T X\, (Q_T^2 \,,\, Q^2 \,,\, y) 
+ \Delta_T Y\, (Q_T^2 \,,\, Q^2 \,,\, y) \right],
\label{cross section}
\end{eqnarray}
where $\Delta_T X$ and $\Delta_T Y$ are, respectively, expressed as 
the convolution of (\ref{tPDF}) with the corresponding partonic cross
sections, see \cite{KKST:06} for their explicit form in the $\overline{\rm MS}$ scheme:
$\Delta_T X = \Delta_T X^{(0)} + \Delta_T X^{(1)}$ as the sum of 
${\cal O}(\alpha_s^0)$ and ${\cal O}(\alpha_s^1)$ contributions, 
where $\alpha_s = \alpha_s(\mu_R^2)$ with $\mu_R$ the renormalization scale, 
and $\Delta_T X^{(0)} = \delta H (x_1^0\,,\,x_2^0\,;\, \mu_F^2 )\ \delta (Q_T^2)$.
The partonic cross section associated with $\Delta_T X^{(1)}$ contains 
all terms that are singular as $Q_T \rightarrow 0$,
behaving $Q_T^{-2} \times (\ln(Q^2 /Q_T^2 )$ or $1)$ or $\delta (Q_T^2)$,
while the ${\cal O}(\alpha_s)$ terms that are less singular than those 
in $\Delta_T X^{(1)}$ are included in the ``finite'' part $\Delta_T Y$. 
In (\ref{cross section}), $\Delta_T X$
becomes very large as $\sim \alpha_s \ln(Q^2/Q_T^2 )/Q_T^2$ and $\sim \alpha_s /Q_T^2$
when $Q_T \ll Q$, representing the recoil effects from the emission 
of the soft and/or collinear gluon, 
and those terms have to be combined with
the large contributions of similar nature that appear in each order of perturbation
theory
as $\alpha_s^n \ln^{2n-1}(Q^2/Q_T^2 )/Q_T^2$, 
$\alpha_s^n \ln^{2n-2}(Q^2/Q_T^2 )/Q_T^2$, and so on, from the multiple gluon emission.

The resummation of those logarithmically enhanced contributions 
to all orders has been worked out \cite{KKST:06}, in order to obtain a
well-defined, finite prediction for the cross section.
This is carried out by exponentiating the soft gluon effects in the 
impact parameter $b$ space, up to the NLL accuracy.
As the result, $\Delta_T X$ of (\ref{cross section})
is replaced by the corresponding NLL resummed component as
$\Delta_T X \rightarrow \Delta_T X^{\rm NLL}$, 
with \cite{KKST:06} 
\begin{equation}
\Delta_T X^{\rm NLL} (Q_T^2 , Q^2 , y) =
\sum_{i,j,k}e_i^2 
\int_0^{\infty} d b \frac{b}{2}
J_0 (b Q_T) e^{S (b , Q)}  
( C_{ij} \otimes f_j )
           \left( x_1^0 , \frac{b_0^2}{b^2} \right)      
 ( C_{\bar{i} k} \otimes f_k )
           \left( x_2^0 , \frac{b_0^2}{b^2} \right).
\label{resum}
\end{equation}
Here $J_0(bQ_T)$ is a Bessel function for the two-dimensional Fourier 
transformation from the $b$ space to the $Q_T$ space, and
$b_0=2e^{-\gamma_E}$ with $\gamma_E$ the Euler constant. 
The symbol $\otimes$ denotes convolution as
$(C_{ij} \otimes f_j )\ (x, \mu^2) = \int_x^1\,
        (d z /z)\, C_{ij} (z, \alpha_s(\mu^2))\, f_j (x / z, \mu^2)$.
Note that the suffix $i , j , k$ can be either $q , \bar{q}$
including the flavor degrees of freedom,
and we set $f_{q}(x, \mu^2) \equiv \delta q ( x , \mu^2)$, 
$f_{\bar{q}}(x, \mu^2) \equiv \delta \bar{q} ( x , \mu^2)$.
The soft gluon effects are resummed into the Sudakov factor $e^{S(b,Q)}$ with 
\bea
S(b,Q)=-\int_{b_0^2/b^2}^{Q^2}\frac{d\kappa^2}{\kappa^2}
\left\{ A_q(\alpha_s(\kappa^2)) \ln \frac{Q^2}{\kappa^2} 
+ B_q (\alpha_s(\kappa^2))\right\}.
\label{sudakov}
\eea
The functions $A_q$, $B_q$ as well as 
the coefficient functions $C_{ij}$ are perturbatively calculable:
$A_q (\alpha_s )= \sum_{n=1}^{\infty} \left( \frac{\alpha_s}{2 \pi}
\right)^n A_q^{(n)}$,  
$B_q (\alpha_s )= \sum_{n=1}^{\infty} \left( \frac{\alpha_s}{2 \pi}
\right)^n B_q^{(n)}$,
and
$C_{ij} (z, \alpha_s )
= \delta_{ij}\delta (1 - z) +
\sum_{n=1}^{\infty} \left( \frac{\alpha_s}{2 \pi}\right)^n C_{ij}^{(n)} (z)$.
At the NLL accuracy, 
\bea
A_q^{(1)}=2C_F,
~~~A_q^{(2)}=2C_F\left\{\left(\frac{67}{18}-\frac{\pi^2}{6}\right)C_G
-\frac{5}{9}N_f\right\}, ~~~B_q^{(1)}=-3C_F, 
\label{eq:AB}
\eea  
where $C_F= (N_c^2 -1)/(2N_c )$, $C_G = N_c$, and $N_f$ is the number of
QCD massless flavors, and 
\bea
C_{ij}^{(1)}(z)
=\delta_{ij} C_F\left(\frac{\pi^2}{2}-4\right)\delta(1-z)
\label{eq:C}
\eea
are derived in \cite{KKST:06}.
The result (\ref{eq:AB}) coincides with that obtained for other
processes \cite{DS:84,KT:82}, demonstrating that $\{A_q^{(1)}, A_q^{(2)}, B_q^{(1)}\}$ 
are universal (process-independent).~\footnote{$B_q^{(n)}$ ($n \geq 2$)
and $C_{ij}^{(n)}(z)$ ($n \geq 1$) depend on the process~\cite{dG}.
Also, $A_q^{(n)}$ ($n=1,2,\ldots$)
and $B_q^{(1)}$ are independent of the factorization
scheme, but $B_q^{(n)}$ ($n \ge 2$) and $C_{ij}^{(n)}(z)$ ($n \ge 1$) 
depend on the factorization scheme
(see e.g. \cite{BCDeG:03}).}
Substituting (\ref{eq:AB}) and 
the running coupling constant $\alpha_s (\kappa^2)$ at two-loop level,
the $\kappa^2$ integral in (\ref{sudakov})
can be performed explicitly to the NLL accuracy, and the result can be
systematically organized as (see also \cite{LKSV:01,BCDeG:03})
\bea
S(b, Q)&=&\frac{1}{\alpha_s (\mu_R^2) }h^{(0)}(\lambda)+h^{(1)}(\lambda)\ ,
\label{sudakov:1}
\eea 
where the first and second terms collect the LL and NLL contributions,
respectively, as 
\bea
h^{(0)}(\lambda)&=&\frac{A_q^{(1)}}{2\pi\beta_0^2}[\lambda+\ln(1-\lambda)],
\label{eq:h0}
\\
h^{(1)}(\lambda)&=&\frac{A_q^{(1)}\beta_1}{2\pi\beta_0^3}
\left[\frac{1}{2}\ln^2(1-\lambda)+\frac{\lambda+\ln(1-\lambda)}{1-\lambda}\right]
+\frac{B_q^{(1)}}{2\pi\beta_0}\ln(1-\lambda)
\nonumber
\\&&
-\frac{1}{4\pi^2\beta_0^2}\left[ A_q^{(2)} - 2\, \pi \beta_0 A_q^{(1)} 
\ln \frac{Q^2}{\mu_R^2} \right]
\left[\frac{\lambda}{1-\lambda}+\ln(1-\lambda)\right].
\label{sudakov:2}
\eea
In these equations, $\beta_0\,,\, \beta_1$ are the first two coefficients
of the QCD $\beta$ function given by
$\beta_0=( 11C_G-2N_f )/(12\pi)$, 
$\beta_1= ( 17C_G^2-5C_GN_f-3C_FN_f)/(24\pi^2 )$, and 
\begin{equation}
\lambda =  \beta_0\alpha_s( \mu_R^2 ) \ln \frac{Q^2 b^2}{b_0^2} 
\equiv \beta_0\alpha_s( \mu_R^2 ) L\ .
\label{eq:lambda}
\end{equation} 
In the $b$ space, $L = \ln(Q^2b^2/b_0^2)$ plays the role of the large logarithmic
expansion parameter with $b\sim 1/Q_T$, and $\lambda$ of
(\ref{eq:lambda}) is formally considered as being of order unity in 
the resummed logarithmic expansion to the NLL in (\ref{sudakov:1}), where 
the neglected NNLL corrections are down by $\alpha_s(\mu_R^2 )$.  
Note that, expanding the above NLL formula (\ref{resum}) with 
(\ref{eq:AB})-(\ref{eq:lambda}) in powers
of $\alpha_s(\mu_R^2)$, the first three towers of logarithms, 
$\alpha_s^n\ln^m(Q^2/Q_T^2)/Q_T^2$ with $m=2n-1,2n-2$
and $2n-3$, in the tDY differential cross section are fully reproduced for all $n$.
Combining this expansion with the finite part $\Delta_T Y$ of (\ref{cross section}), 
the result gives the tDY differential cross section which is exact up to 
${\cal O}(\alpha_s)$; 
thus we use the NLO parton distributions in the $\overline{\rm MS}$ scheme for
$f_j (x, \mu^2)$ in (\ref{resum}), as well as for those involved in $\Delta_T Y$. 

We explain some further manipulations for our NLL formula; those 
were actually performed in \cite{KKST:06}, but were not described in detail.
The integrand of (\ref{resum})
depends on the parton distributions at the scale $b_0/b$, according to the general 
formulation \cite{CSS:85}. 
Taking the Mellin moments of $\Delta_T X^{\rm NLL} (Q_T^2 , Q^2 , y)$ 
with respect to the DY scaling variables $x_{1,2}^{0}$ at fixed $Q$, 
\begin{equation}
\Delta_T X^{\rm NLL}_{N_1, N_2}  (Q_T^2 , Q^2)
\equiv \int_{0}^1 dx_{1}^0 \left(x_1^{0} \right)^{N_1 -1}  \int_{0}^1 dx_2^{0} 
\left(x_2^{0} \right)^{N_2 -1} \Delta_T X^{\rm NLL} (Q_T^2 , Q^2 , y),
\label{eq:MT}
\end{equation}
the $b$-dependence of those parton distributions can be disentangled because 
the moments, $f_{i,N}(\mu^2 )\equiv \int_0^1 dx x^{N-1} f_{i}(x, \mu^2 )$,
obey the renormalization group (RG) evolution as
$f_{i,N} (b_0^2 /b^2 )$ $=\sum_{j} U_{ij,N} (b_0^2 /b^2 , Q^2 ) f_{j,N}(Q^2)$,
where $U_{ij,N}(\mu^2 , {\mu'}^2 )$ are the NLO evolution operators for 
the transversity distributions which are expressed in terms of 
the corresponding LO and NLO anomalous dimensions \cite{AM:90,KMHKKV:97} 
and the two-loop running coupling constant. 
For (\ref{eq:MT}) with (\ref{resum}) and the above RG evolution 
substituted, 
several ``reorganization'' of the
relevant large-logarithmic expansion is necessary for its consistent evaluation
over the entire range of $Q_T$, 
following the systematic procedure in \cite{BCDeG:03} elaborated for
unpolarized hadron collisions:
exploiting the RG invariance,
we have $C_{ij,N} (\alpha_s(b_0^2 / b^2 )) = C_{ij,N} (\alpha_s( Q^2))
e^{[\alpha_s(\mu_R^2 )C_{ij,N}^{(1)} /2\pi] \lambda/(1-\lambda)}$ 
to the corrections down by $\alpha_s(\mu_R^2 )$, 
for the $N$-th moment 
of the coefficient function of (\ref{eq:C}),
so that we make the replacement 
$C_{ij,N} (\alpha_s(b_0^2 / b^2 )) \rightarrow C_{ij,N} (\alpha_s( Q^2))
= \delta_{ij}[1+ (\alpha_s (Q^2 ) C_F/4 \pi) (\pi^2 -8) ]$, 
up to the corrections of NNLL level for (\ref{eq:MT}).
Similarly, 
performing the large-logarithmic expansion for explicit formula 
of the NLO evolution operator $U_{ij,N} (b_0^2 /b^2 , Q^2 )$,
we find
\bea
U_{ij,N}(b_0^2 /b^2 , Q^2 ) = \delta_{ij}e^{R_N(\lambda)}, \;\;\;\;\;\;\;\;
R_{N}(\lambda)\equiv \frac{\Delta_T P_{qq,N}}{2\pi\beta_0}\ln(1-\lambda),
\label{LO-evol}
\eea
up to the corrections down by $\alpha_s(\mu_R^2 )$ which correspond to
the NNLL terms when substituted into (\ref{eq:MT}), (\ref{resum}).
Here $\Delta_T P_{qq,N} = -2C_F [\psi(N+1) + \gamma_E - 3/4 ]$ is the 
$N$-th Mellin moment of the LO DGLAP splitting function for the transversity.
As a result, (\ref{eq:MT}) is expressed as
\bea
\Delta_T X^{\rm NLL}_{N_1, N_2}  (Q_T^2 , Q^2)
&&=
\left[1+\frac{\alpha_s (Q^2)}{2\pi}C_F(\pi^2-8) \right]
\delta H_{N_1,N_2}(Q^2)
I_{N_1,N_2}(Q_T^2, Q^2)\ ,
\label{resum:2} \\
I_{N_1,N_2}(Q_T^2, Q^2) && \equiv 
\int_0^{\infty} d b \frac{b}{2} J_0 (b Q_T)
e^{S (b, Q)+ R_{N_1}(\lambda)+R_{N_2}(\lambda)}\ ,
\label{resum:21}
\eea 
where $\delta H_{N_1,N_2}(Q^2)$ is the double Mellin-moments of 
$\delta H(x_1^0, x_2^0 ;  Q^2)$ of (\ref{tPDF}),
defined similarly as (\ref{eq:MT}). The complete dependence on $b$
is included in the exponential factor $e^{S (b, Q)+ R_{N_1}(\lambda)+R_{N_2}(\lambda)}$ 
through $L = \ln(Q^2b^2/b_0^2)$,
so that all-order resummation of the large logarithms $L$
and the associated $b$-integral in (\ref{resum:21})
are now accomplished at the partonic level. 

We also mention some other ``reorganization'', 
which is explained in \cite{KKST:06} and 
is necessary 
in order to treat properly too short and long distance involved 
in the $b$ integration of (\ref{resum:21}):
firstly, to treat too short distance $Qb \ll 1$,
we make the replacement 
\begin{equation}
L\ra\tilde{L}=\ln(Q^2b^2/b_0^2+1)\ , 
\label{replaceL}
\end{equation}
in the definition (\ref{eq:lambda}) of $\lambda$,
following \cite{BCDeG:03};
note that the integrand of (\ref{resum:21})
depends on the large-logarithmic expansion 
parameter only through $\lambda$ 
(see (\ref{sudakov:1})-(\ref{sudakov:2}), (\ref{LO-evol})).
This replacement allows us to reduce 
the unjustified large logarithmic contributions for $Qb \ll 1$,
due to $L \gg 1$, as $\tilde{L} \rightarrow 0$ and 
$e^{S(b,Q)+R_{N_1}(\lambda)+R_{N_2}(\lambda)} \rightarrow 1$, 
while 
$L$ and $\tilde{L}$ are equivalent to organize 
the soft gluon resummation at small $Q_T$ as
$\tilde{L}=L+{\cal O}(1/(Qb)^2 )$ for $Qb \gg 1$.
Secondly, the functions 
(\ref{eq:h0}) and (\ref{sudakov:2}) 
in the Sudakov exponent (\ref{sudakov:1}) are singular when 
$\lambda = \beta_0\alpha_s( \mu_R^2 ) \tilde{L} \rightarrow 1$, 
and this singular behavior 
is related to the presence of the Landau pole 
in the perturbative running coupling $\alpha_s (\kappa^2)$
in QCD. To properly define the $b$ integration of (\ref{resum:21}) 
for the corresponding long-distance region,
it is necessary to specify a prescription to deal with
this singularity
\cite{LKSV:01}:
decomposing the Bessel function in (\ref{resum:21}) into the two Hankel functions as
$J_0(bQ_T) = (H_0^{(1)}(bQ_T )+H_0^{(2)}(bQ_T ) )/2$,
we deform the $b$-integration contour for these two terms 
into upper and lower half plane in the complex $b$ space, respectively, 
and obtain the two convergent integrals as $|b| \rightarrow \infty$.
The new contour ${\cal C}$ is taken as:
from $b= 0$ to $b=b_c$ on the real axis, 
followed by the two branches,
$b=b_c + e^{\pm i\theta}t$ with $t \in \{0, \infty \}$ and $0<\theta<\pi/4$;
a constant $b_c$ is chosen as $0 \le b_c < b_L$, 
where $b=b_L$ gives the solution for $\lambda =1$.
Note, this choice of contours 
is completely equivalent to the original contour,
order-by-order in $ \alpha_s(\mu_R^2 )$, when the corresponding formulae 
are expanded in powers
of $\alpha_s$. Therefore, 
this contour deformation prescription provides us with 
a (formally) consistent definition of finite $b$-integral of (\ref{resum:21})
within a perturbative framework. 

We now denote (\ref{resum:2}), with the replacement (\ref{replaceL})
and the new contour ${\cal C}$ in (\ref{resum:21}),
as $\Delta_T \tilde{X}^{\rm NLL}_{N_1, N_2}  (Q_T^2 , Q^2)$, and also denote 
the double inverse Mellin transform of 
$\Delta_T \tilde{X}^{\rm NLL}_{N_1, N_2}  (Q_T^2 , Q^2)$,
from $(N_1 , N_2 )$ space to $(x_1^0 , x_2^0 )$ space,
as $\Delta_T \tilde{X}^{\rm NLL}  (Q_T^2 , Q^2, y)$.
Defining  (see (\ref{cross section}))
\begin{equation}
\Delta_T  \tilde{Y} (Q_T^2 , Q^2, y) \equiv \Delta_T   X (Q_T^2 , Q^2, y) 
+ \Delta_T   Y (Q_T^2 , Q^2, y) 
- \left. \Delta_T \tilde{X}^{\rm NLL}  (Q_T^2 , Q^2, y) \right|_{\rm FO}\ ,
\label{matching}
\end{equation}
where $\Delta_T \tilde{X}^{\rm NLL}  (Q_T^2 , Q^2, y) |_{\rm FO}$ denotes the terms
resulting from the expansion of the resummed expression up to the
fixed-order $\alpha_s(\mu_R^2 )$,
we obtain the final form of our differential cross section 
for tDY with the soft gluon resummation as \cite{KKST:06} 
\bea
\frac{\Delta_Td\sigma}{dQ^2dQ_T^2dyd\phi}=
\cos(2\phi)
\frac{\alpha^2}{3\, N_c\, S\, Q^2}
\biggl[\Delta_T\tilde{X}^{\rm NLL}(Q_T^2 , Q^2,y)
+\Delta_T\tilde{Y}(Q_T^2 , Q^2,y)\biggr].
\label{NLL+LO}
\eea
From the derivation explained above, the expansion of 
this cross section in powers of $\alpha_s(\mu_R^2 )$
fully reproduces the first three towers of logarithms, 
$\alpha_s^n\ln^m(Q^2/Q_T^2)/Q_T^2$ with $m=2n-1,2n-2$
and $2n-3$, associated with the soft-gluon emission for small $Q_T$ ($\ll Q$), 
and also coincides exactly with the fixed-order result
(\ref{cross section}) to ${\cal O}(\alpha_s)$. Therefore, this formula (\ref{NLL+LO}) 
avoids any double counting
over the entire range of $Q_T$. Note that $\Delta_T  \tilde{Y} (Q_T^2 , Q^2, y)$ 
of (\ref{matching}) corresponds to 
the ``modified finite component'' in our resummation framework:
because the first and the third terms in the RHS of (\ref{matching}) cancel with each
other for  $Q_T \ll Q$, $\Delta_T  \tilde{Y} (Q_T^2 , Q^2, y)$ 
is less singular as $Q_T \rightarrow 0$ than $Q_T^{-2} \times (\ln(Q^2 /Q_T^2 )$ or $1)$ 
or $\delta (Q_T^2)$, see the discussion below (\ref{cross section}). Combined with 
$\Delta_T  X^{(0)} \propto \delta (Q_T^2)$, this also implies that  
$\Delta_T  \tilde{Y} (Q_T^2 , Q^2, y)$ is of order $\alpha_s (\mu_R^2 )$.
In fact, (\ref{matching}) coincides exactly with $\Delta_T  Y  (Q_T^2 , Q^2, y)$  
if (\ref{replaceL}) is not performed.

Because of this ``regular'' behavior of $\Delta_T  \tilde{Y} (Q_T^2 , Q^2, y)$ 
as $Q_T \rightarrow 0$,
we may consider (\ref{matching}) as the definition for the region where $Q_T > 0$;
in this case, the first two terms correspond to (\ref{cross section})
for $Q_T > 0$, i.e.,
\begin{eqnarray}
\frac{\Delta_T d \sigma^{\rm LO}}{d Q^2 d Q_T^2 d y d \phi}
= 
\cos{(2 \phi )}
\frac{\alpha^2}{3\, N_c\, S\, Q^2}
 \left[ \Delta_T \left. X^{(1)}\, (Q_T^2 , Q^2 , y)\right|_{Q_T^2 >0} 
+ \Delta_T Y\, (Q_T^2 , Q^2 , y) \right],
\label{LOcross section}
\end{eqnarray}
which gives the formula for the LO QCD prediction of tDY at the
large-$Q_T$ region. Therefore, our formula (\ref{NLL+LO})
is actually the NLL resummed part, with the contributions to ${\cal O}(\alpha_s )$  
(the third term of (\ref{matching})) subtracted, plus the LO cross section;
we refer to (\ref{NLL+LO}) as the ``NLL+LO'' prediction, 
which gives the well-defined tDY differential cross section 
in the $\overline{\rm MS}$ scheme over the entire range of $Q_T$.  
It is straightforward to see that the integral of (\ref{NLL+LO}) over $Q_T$ reproduces
that of (\ref{cross section}) exactly, because
$\tilde{L}= 0$ at $b=0$ (see also \cite{BCDeG:03}).
 
We can extend the above results to unpolarized DY by mostly trivial substitutions
to switch from spin-dependent quantities to spin-averaged ones,
e.g., by removing ``$\Delta_T$'' 
and making the replacement, $\delta H(x_1, x_2 ; \mu^2) 
\rightarrow H(x_1, x_2 ; \mu^2)$,
$\cos(2\phi) \alpha^2 / (3 N_c S  Q^2)$ $\rightarrow 2 \alpha^2 / (3 N_c S  Q^2)$, etc.,
in the above relevant formulae.
The explicit form of the spin-averaged quantities,
such as $X (Q_T^2 , Q^2, y)$, $Y (Q_T^2 , Q^2, y)$, as well as those 
corresponding to the coefficient functions $C_{ij}(z, \alpha_s )$ in (\ref{resum}),
can be obtained from the results in \cite{CSS:85,AEGM:84}.
A different point from the polarized case is that now the gluon distribution
$f_g(x, \mu^2) \equiv g(x, \mu^2 )$ participates,
so that the suffix $i,j$ of the ``spin-averaged $C_{ij}(z, \alpha_s )$''
can be ``$g$'' as well as ``$q,\bar{q}$''.
This also implies that $\Delta_T P_{qq,N}$ appearing in (\ref{LO-evol}) has to
be replaced by the Mellin moment of the 
LO DGLAP splitting functions for the unpolarized case,
which involve the mixing of gluon, and the ``new $U_{ij,N}(b_0^2 /b^2 , Q^2)$''
represent the corresponding ``evolution matrix'' that was discussed in 
\cite{BCDeG:03,LKSV:01}.
On the other hand, the formulae (\ref{sudakov:1})-(\ref{sudakov:2}) of the 
Sudakov exponent hold also for the unpolarized case,
reflecting that the coefficients (\ref{eq:AB}) relevant at the NLL level
are universal \cite{KT:82,dG,LKSV:01,BCDeG:03}.
We list explicit form of the relevant formulae for the unpolarized cross sections 
in Appendix.

Taking the ratio of (\ref{NLL+LO}) to the corresponding NLL+LO prediction for 
unpolarized differential cross section, we obtain the 
double transverse-spin asymmetry in tDY, for transverse-momentum $Q_T$, 
invariant-mass $Q$, 
and rapidity $y$ of the produced lepton pair, and azimuthal angle $\phi$
of one of the leptons, as
\bea
\aqt=\frac{1}{2} \cos(2\phi) 
\frac{\Delta_T\tilde{X}^{\rm NLL}(Q_T^2 , Q^2, y)
+\Delta_T\tilde{Y}(Q_T^2, Q^2, y)}
{\tilde{X}^{\rm NLL}(Q_T^2, Q^2, y)+\tilde{Y}(Q_T^2, Q^2, y)}.
\label{asym}
\eea
To the fixed-order $\alpha_s$ without the soft gluon resummation, 
(\ref{asym}) reduces to the LO prediction of the asymmetry for $Q_T >0$,
\bea
\aqtn^{\rm LO}(Q_T)
=\frac{1}{2} \cos(2\phi) 
\frac{\left. \Delta_T X^{(1)}(Q_T^2 , Q^2, y) \right|_{Q_T^2 >0}
+\Delta_T Y(Q_T^2, Q^2, y)}
{\left. X^{(1)}(Q_T^2, Q^2, y)\right|_{Q_T^2 >0} + Y(Q_T^2, Q^2, y)},
\label{asymlo}
\eea
as the ratio of (\ref{LOcross section}) to the corresponding unpolarized cross section.

\section{The asymmetries $\aqt$ at RHIC and J-PARC}

We evaluate the asymmetries, derived in the last section, as a function of $Q_T$. 
We use the similar parton distributions as in the previous NLO studies
\cite{MSSV:98} of $Q_T$-independent $A_{TT}$ of (\ref{eq:att}):
for the transversity $\delta q(x, Q^2)$ participating in the numerator 
of the asymmetries, we use a model of the NLO transversity
distributions, 
which obey the corresponding NLO DGLAP evolution equation
and are assumed to saturate the 
Soffer bound \cite{Soffer:95} 
as $\delta q(x,\mu^2_0)=[q(x,\mu^2_0)+\Delta q(x,\mu^2_0)]/2$  
at a low input scale $\mu_0\simeq 0.6$ GeV 
using the NLO GRV98 \cite{GRV:98} and GRSV2000 (``standard scenario'') \cite{GRSV:00} 
distributions $q(x,\mu_0^2)$ 
and $\Delta q(x,\mu^2_0)$, respectively.
The NLO GRV98 distributions $q(x, Q^2), g(x, Q^2)$ are also used for calculating 
the unpolarized cross sections in the denominator of the asymmetries.

It is known that the $Q_T$-spectrum of DY lepton pair is affected 
by another nonperturbative effects,
which become important for small $Q_T$ region \cite{CSS:85}:
we have obtained the well-defined tDY cross sections and asymmetries
that are free from any singularities, 
with a consistent definition of the integration in (\ref{resum:21})
over the whole $b$ region.
However, the integrand of (\ref{resum:21}) involving purely perturbative quantities
is not accurate for extremely large $|b|$ region in QCD, and 
the corresponding long-distance behavior
has to be complemented by the relevant nonperturbative effects.
Formally those nonperturbative effects play role 
to compensate the ambiguity that the prescription for the $b$ 
integration in (\ref{resum:21}) to avoid the singularity
in the Sudakov exponent $S(b, Q)$ of (\ref{sudakov:1})-(\ref{sudakov:2})
is actually not unique (see \cite{CSS:85}). Therefore, 
following \cite{CSS:85,LKSV:01,BCDeG:03}, we make the replacement in (\ref{resum:21}) as 
\begin{equation}
e^{S (b , Q)}\rightarrow e^{S (b , Q)-  g_{NP} b^2} ,
\label{eq:np}
\end{equation}
with a nonperturbative parameter $g_{NP}$.
Because exactly the same Sudakov factor $e^{S(b, Q)}$ 
participates in the corresponding formula for the unpolarized case 
as noted above (\ref{asym}), we perform the
replacement (\ref{eq:np}) with the same nonperturbative parameter $g_{NP}$
in the NLL+LO unpolarized differential cross section contributing to the denominator of
(\ref{asym}).
This may be interpreted as assuming the same ``intrinsic transverse
momentum'' of partons inside nucleon for both polarized and unpolarized cases, 
corresponding to the Gaussian smearing factor of (\ref{eq:np}).
We use $g_{NP}\simeq 0.5$ GeV$^2$, 
suggested by the study of the $Q_T$-spectrum in unpolarized case \cite{KS:03}.

For all the following numerical evaluations, we choose 
$\phi=0$ for the azimuthal angle of one lepton, $\mu_F =\mu_R =Q$
for the factorization and renormalization scales and 
$b_c=0$, $\theta=\frac{7}{32}\pi$ for the integration contour 
${\cal C}$ explained below (\ref{replaceL}). 

\begin{figure}
\bc
\includegraphics[height=5.8cm]{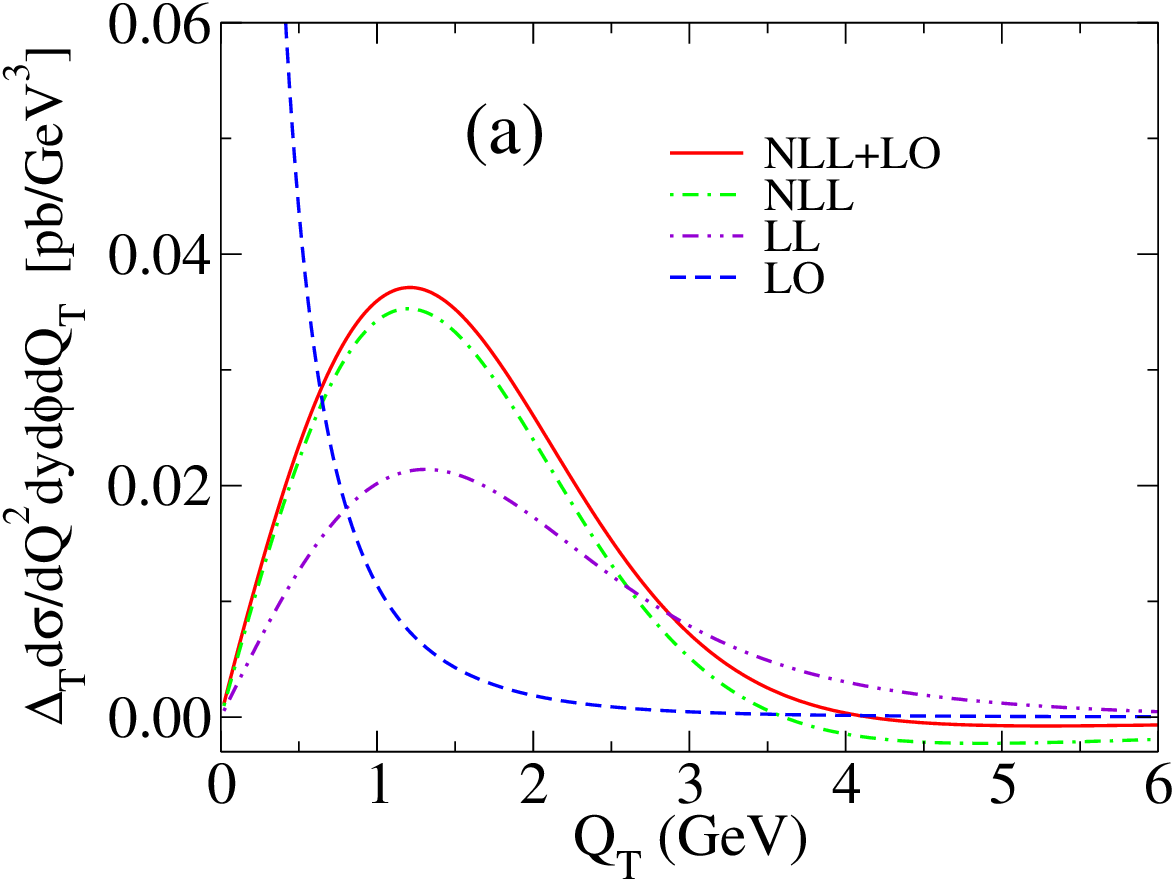}~~~~
\includegraphics[height=5.8cm]{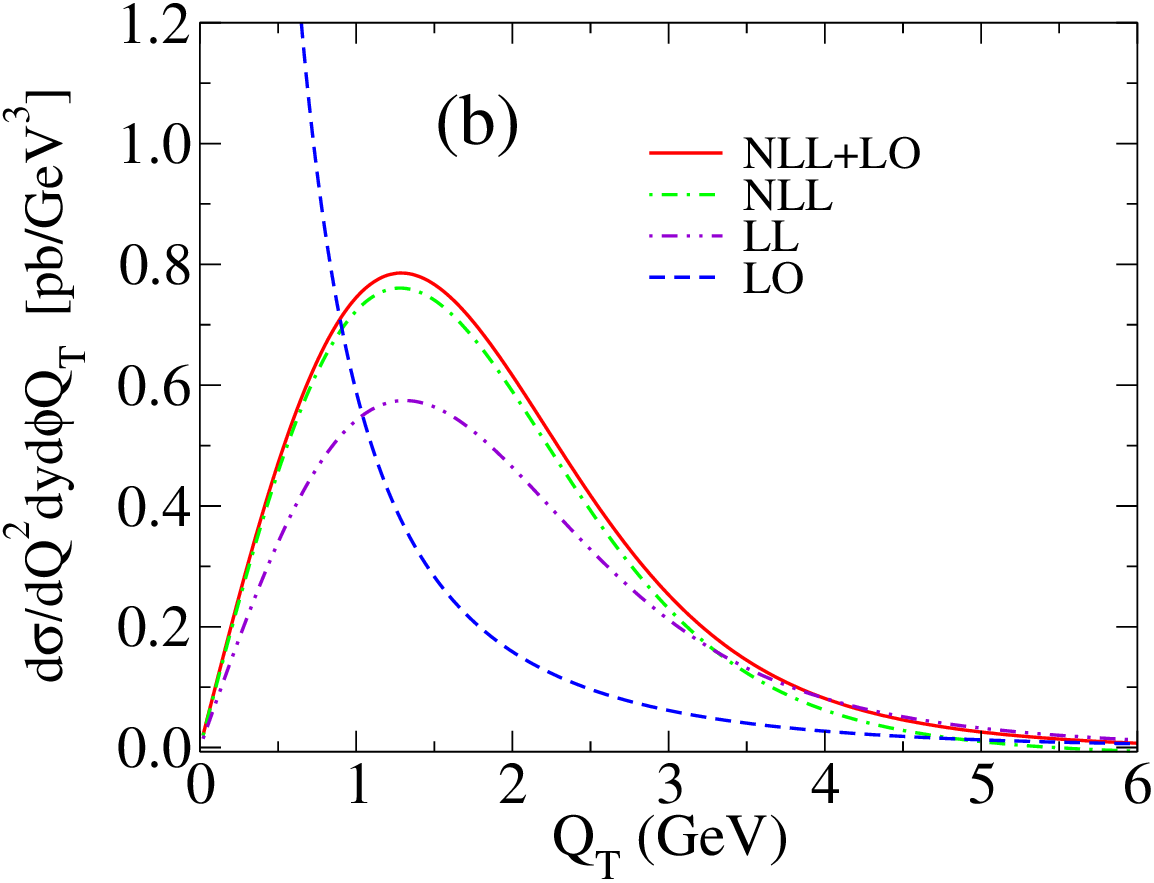}
\ec
\caption{The spin-dependent and spin-averaged differential 
cross sections for tDY: (a) $\Delta_Td\sigma/dQ^2 dQ_T dy d\phi$ 
and 
(b) $d\sigma/dQ^2 dQ_T dyd\phi$,
as a function of $Q_T$ 
at RHIC kinematics, $\sqrt{S}=200$ GeV, $Q=5$ GeV, $y=2$ and $\phi=0$,
with $g_{NP}=0.5$ GeV$^2$. 
}
\label{fig:1}
\end{figure}

First of all,
we present the transvserse-momentum $Q_T$-spectrum of the DY lepton pair 
for $\sqrt{S}=200$~GeV, $Q=5$~GeV, and $y=2$,
which correspond to the detection of dileptons with the PHENIX detector at RHIC.
The solid curve in Fig.~\ref{fig:1}(a)
shows the NLL+LO differential cross section (\ref{NLL+LO}) for tDY, multiplied by $2Q_T$,
with $g_{NP}=0.5$ GeV$^2$ for (\ref{eq:np}).
We also show the contribution from the NLL resummed component 
$\Delta_T \tilde{X}^{\rm NLL}$ in (\ref{NLL+LO}) by the dot-dashed curve, and
the LO result using (\ref{LOcross section}) by the dashed curve.
Fig.~\ref{fig:1}(b) is same as Fig.~\ref{fig:1}(a)
but for the unpolarized differential cross sections.
The LO results become large and diverge as $Q_T \rightarrow 0$,
while the NLL+LO results are finite and well-behaved over all regions of $Q_T$. 
The soft gluon resummation
gives dominant contribution around the peak of the solid curve,
i.e., 
at intermediate $Q_T$ as well as small $Q_T$.
To demonstrate the resummation effects in detail,
the two-dot-dashed curves in Figs.~\ref{fig:1}(a), (b) show the LL result which 
is obtained from the corresponding NLL result (dot-dashed curve)
by omitting the contributions corresponding to the NLL level, 
i.e., $h^{(1)}(\lambda)$, $R_{N_1}(\lambda)$, $R_{N_2}(\lambda)$ 
in (\ref{resum:21}) and $\alpha_s(Q^2) C_F (\pi^2 -8 )/2\pi$
in (\ref{resum:2})
for the polarized case (see (\ref{sudakov:1}) and the discussion below
(\ref{eq:lambda})), and similarly for the unpolarized case.
The LL contributions are sufficient for obtaining the finite cross section,
causing considerable suppression in the small $Q_T$ region.
On the other hand, it is remarkable that the contributions at the NLL level
provide significant enhancement from the LL result,
around the peak region for both polarized and unpolarized cases,
and the effect is more pronounced for the former.
Among the relevant NLL contributions, 
the ``universal'' term $h^{(1)}(\lambda)$ 
produces similar (enhancement) effect for both 
(a) and (b) of Fig.~\ref{fig:1}, while the other NLL contributions, associated 
with the evolution operators and the ${\cal O}(\alpha_s(Q^2) )$ coefficient 
functions (see e.g. (\ref{LO-evol}), (\ref{resum:2})),
give different effects to the polarized and unpolarized cases.

\begin{figure}
\bc
\includegraphics[height=5.8cm]{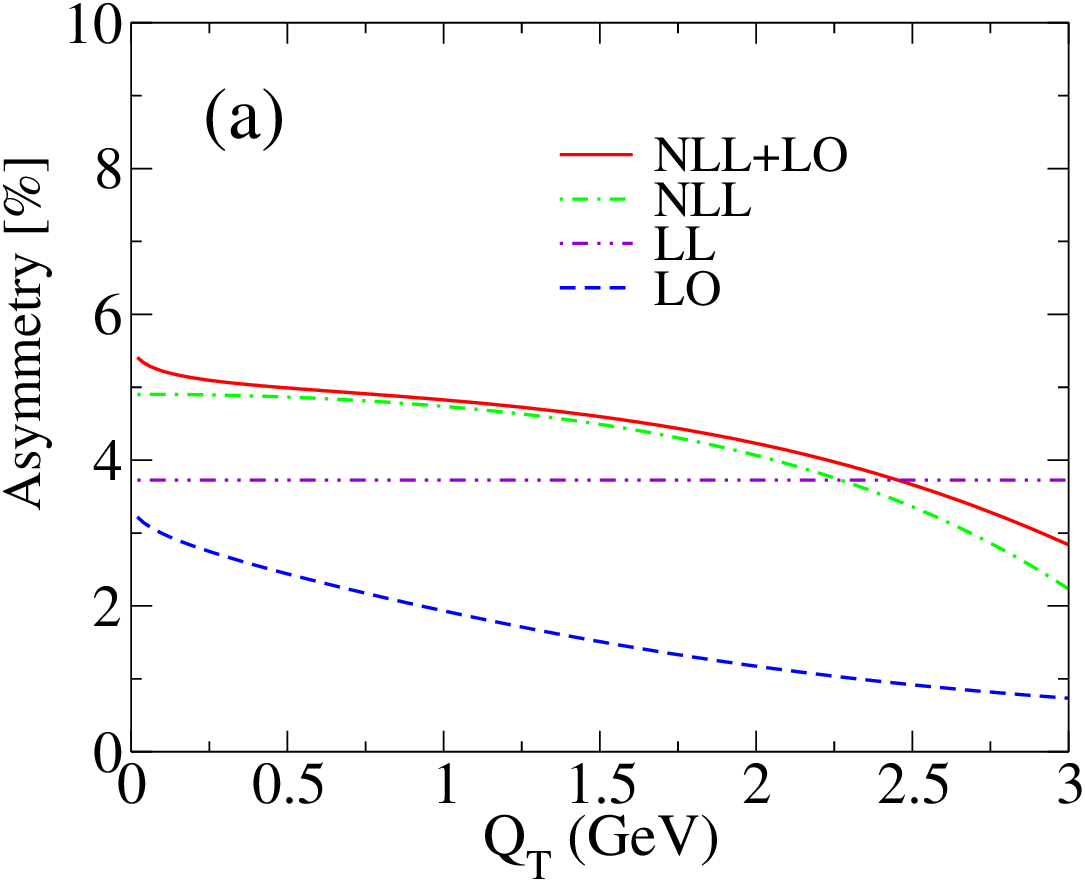}~~~~~~~~
\includegraphics[height=5.8cm]{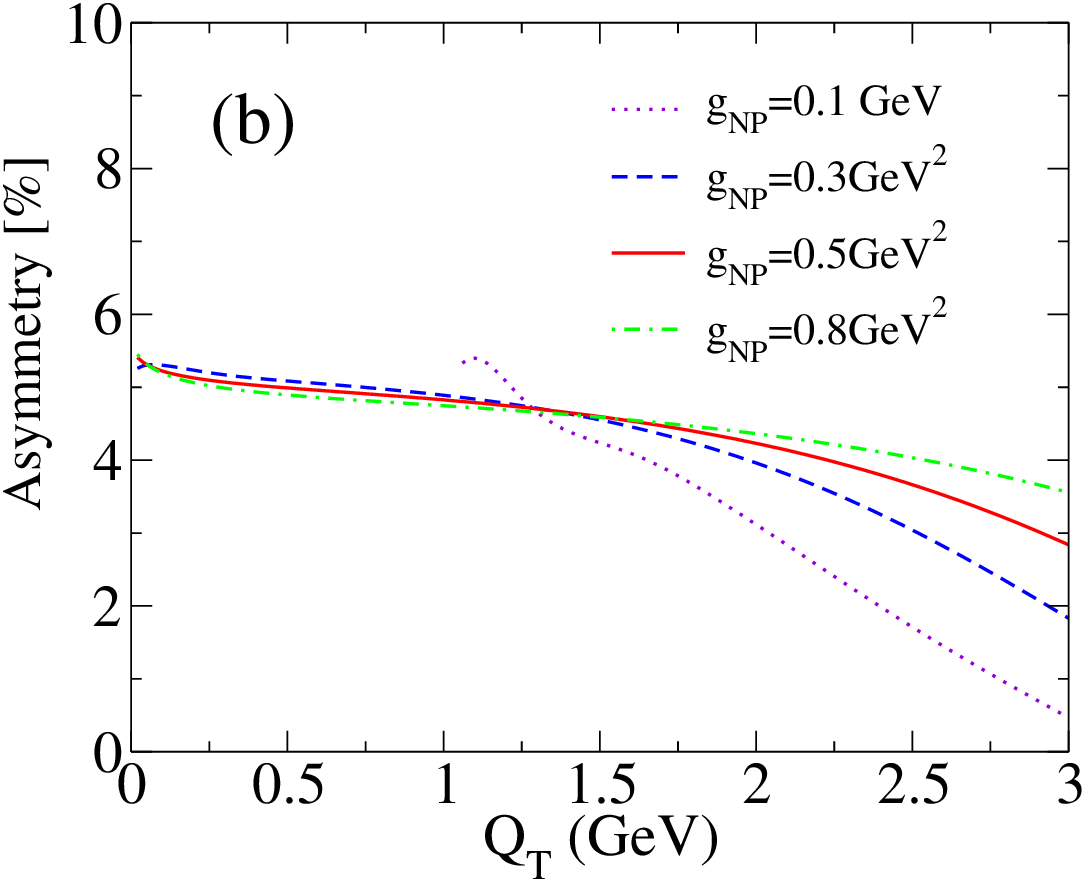}
\ec
\caption{The asymmetries $\aqt$ at RHIC kinematics,
$\sqrt{S}=200$ GeV, $Q=5$ GeV, $y=2$ and $\phi=0$:
(a) $\aqt$ obtained from each curve in Fig.~\ref{fig:1}.
(b) The NLL+LO $\aqt$ of (\ref{asym}) with (\ref{eq:np}) using various values
for $g_{NP}$.
}
\label{fig:2}
\end{figure}

Fig.~\ref{fig:2}(a) shows the double transverse-spin asymmetries 
in the small $Q_T$ region for tDY
at RHIC, obtained as the ratio of the results in 
Fig.~\ref{fig:1}(a) to
the corresponding results in Fig.~\ref{fig:1}(b) 
for respective lines,
so that the solid curve gives the NLL+LO result (\ref{asym}),
the dot-dashed curve shows the NLL result,
\begin{equation}
\aqtn^{\rm NLL}(Q_T) = \frac{1}{2}\cos(2\phi)
\frac{\Delta_T \tilde{X}^{\rm NLL} (Q_T^2 , Q^2 ,y)}{\tilde{X}^{\rm NLL} 
(Q_T^2 , Q^2 ,y)}\ ,
\label{asymNLL}
\end{equation}
and the dashed curve shows the LO result (\ref{asymlo}).
The NLL+LO result 
is almost flat for $Q_T \rightarrow 0$ as well as around the peak region 
of the NLL+LO cross section in Fig.~\ref{fig:1}.
This flat behavior is dominated by the NLL resummed components,
and reflects the fact that the soft gluon emission effects resummed 
into the Sudakov factor $e^{S(b,Q)}$ with (\ref{sudakov:1}) are
universal to the NLL accuracy between the numarator and denominator of (\ref{asym}).
Slight increase of the solid line for $Q_T \rightarrow 0$ is due to the terms 
$\propto \ln (Q^2 /Q_T^2)$ contained in the ``regular components'' 
$\Delta_T \tilde{Y}$ and $\tilde{Y}$ in (\ref{asym}) (see
(\ref{matching})), 
but such weak singularities which show up only at very small $Q_T$ will
be irrelevant for most practical purposes.
The LO result, obtained as the ratio of the two LO curves divergent as 
$Q_T \rightarrow 0$ in Figs.~\ref{fig:1}(a) and (b),
gives the finite asymmetry for $Q_T >0$, but it does not have the flat behavior, i.e., 
decreases for increasing $Q_T$, and is much smaller
than the NLL+LO result. On the other hand, we note that the LL result, retaining only
the resummmed components corresponding to the LL level, is given by 
(see (\ref{resum:2}), (\ref{resum:21}))
\begin{equation}
\aqtn^{\rm LL}(Q_T)=\frac{1}{2}\cos(2 \phi) 
\frac{\delta H(x_1^0, x_2^0; Q^2)}{H(x_1^0,x_2^0; Q^2)} \approx A_{TT}\ ,
\label{LL}
\end{equation}
which is independent of $Q_T$, because the $Q_T$-dependent factor (\ref{resum:21})
with $S(b, Q) + R_{N_1}(\lambda) +R_{N_2}(\lambda) 
\rightarrow h^{(0)}(\lambda)/ \alpha_s(Q^2)$ 
is common for both polarized and unpolarized cases.
Namely the LL resummation effects cancel exactly  
between the numerator and the denominator in the asymmetry (\ref{LL}).
As indicated in (\ref{LL}), the resulting value shown by the two-dot-dashed curve in 
Fig.~\ref{fig:2}(a) coincides with the $Q_T$-independent asymmetry (\ref{eq:att})
up to the NLO QCD corrections; note that $A_{TT}=4.0$\% including the NLO corrections 
similarly as \cite{MSSV:98} (see Table \ref{tab:1} below).
However, 
we recognize
that the soft-gluon resummation contributions 
at the NLL level enhances 
the asymmetry at the small $Q_T$-region significantly, compared with the
LL or fixed-order result.
This is caused by the enhancement of the cross sections in
Fig.~\ref{fig:1} discussed above,
due to the universal $h^{(1)}(\lambda)$ term and the other spin-dependent contributions.
In particular, the evolution operators like (\ref{LO-evol})
in the latter contributions
allow the participation of the parton distributions at the scale $b_0 /b \sim Q_T$,
and the components associated with those parton distributions
indeed play dominant roles due to the mechanism embodied 
by the Sudakov factor $e^{S(b,Q)}$ of (\ref{resum:21}).
Combined with the different $x$-dependence between the transversity and density 
distributions as noted in (ii) above,
the resulting enhancement arises differently 
between (a) and (b) in Fig.~\ref{fig:1}, and
enhances the asymmetry as in Fig.~\ref{fig:2}(a). 

In Fig.~\ref{fig:2}(b) we show the NLL+LO asymmetries $\aqt$ of (\ref{asym}),
with (\ref{eq:np})
using various values of $g_{NP}$.
Here the solid curve is the same as the solid curve 
in Fig.~\ref{fig:2}(a), using $g_{NP}=0.5$ GeV$^2$.
The result demonstrates that our NLL+LO asymmetry in the relevant small-$Q_T$ region
is almost independent of the value of $g_{NP}$ in the range $g_{NP}=0.3$-0.8 GeV$^2$.
Although, at RHIC kinematics, the $Q_T$-spectrum 
from the spin-dependent cross section (\ref{NLL+LO}) with (\ref{eq:np})
receives a sizable smearing effect in the relevant small-$Q_T$ region \cite{KKST:06}, 
the corresponding $g_{NP}$-dependence
is canceled by the similar dependence of the unpolarized cross section
in the asymmetry (\ref{asym}).
In our framework, such cancellation of the
$g_{NP}$-dependence between the numerator and the denominator of (\ref{asym})
is observed for all relevant kinematics of our interest at RHIC, 
and also at J-PARC discussed below.
However, we mention that too small value of $g_{NP}$ is useless in  practice:
the Gaussian smearing factor of (\ref{eq:np}) for $g_{NP}=0.1$ GeV$^2$ 
is insufficient to suppress sensitivity to the extremely large $|b|$
region in (\ref{resum:21}), 
so that the $b$ integration 
receives the ``inaccurate'' long-distance perturbative contributions
considerably at small $Q_T$,
which lead to unstable numerical behavior for $Q_T \lsim 1$~GeV.
For all the following calculations, we use $g_{NP}=0.5$ GeV$^2$.
\begin{figure}
\bc
\includegraphics[height=5.8cm]{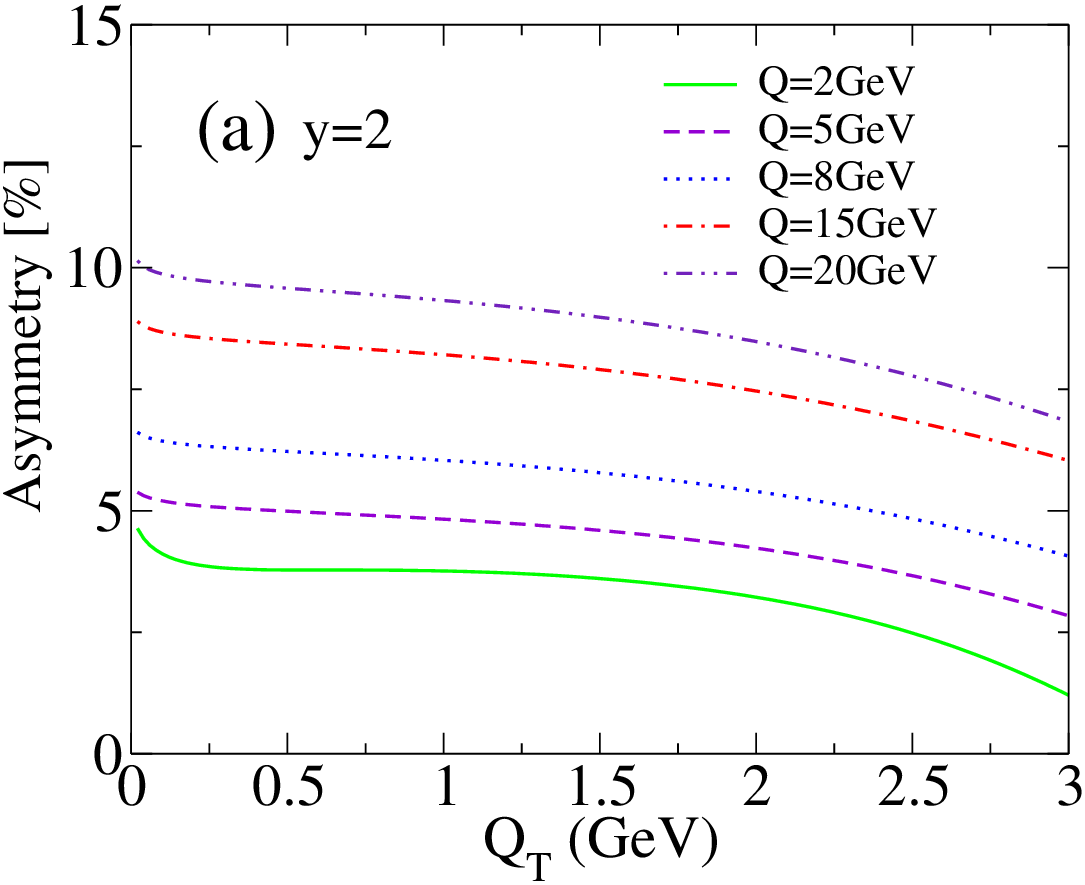}~~~~~~~
\includegraphics[height=5.8cm]{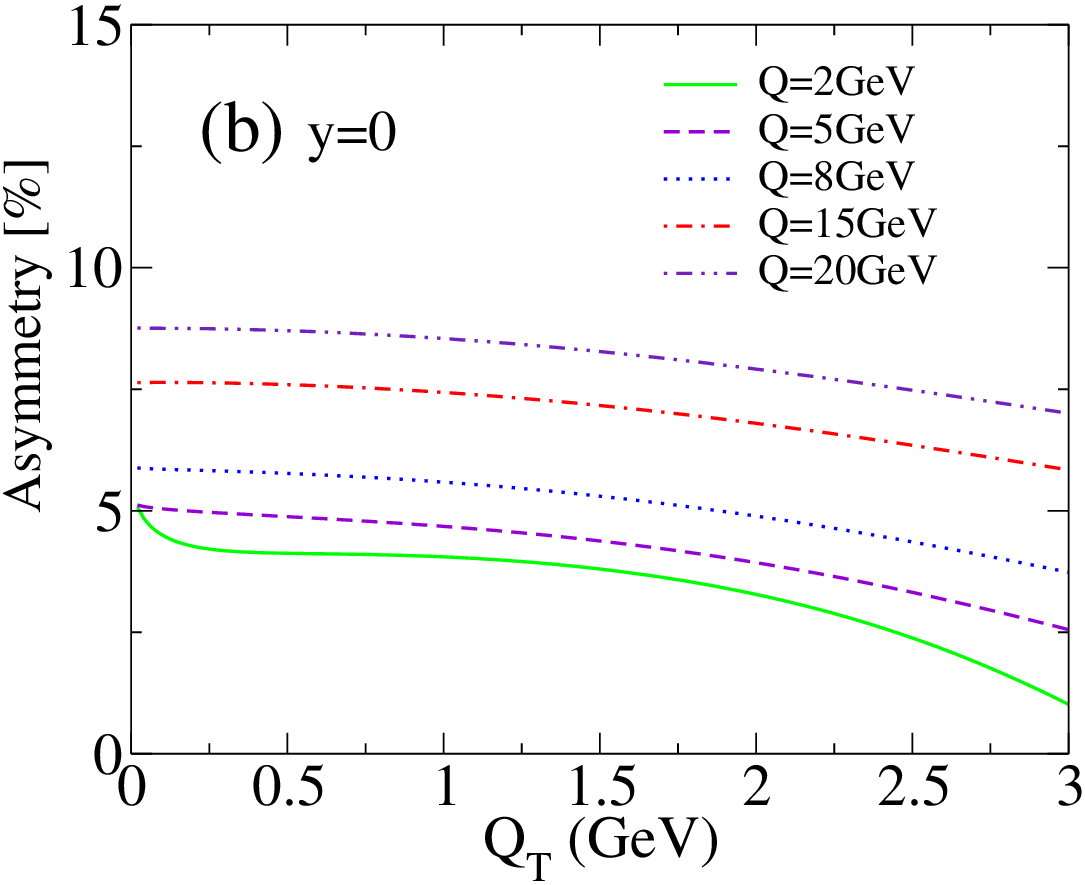} 
\ec
\caption{The NLL+LO $\aqt$ of (\ref{asym}) with (\ref{eq:np}) using $g_{NP}=0.5$ GeV$^2$
at RHIC kinematics, $\sqrt{S}=200$ GeV, $\phi=0$ with $y=2$ and $y=0$ for (a) and (b),
respectively.}
\label{fig:3}
\end{figure}

Fig.~\ref{fig:3} shows the NLL+LO asymmetries 
$\aqt$ of (\ref{asym}) at RHIC kinematics,
$\sqrt{S}=200$ GeV and various values of the dilepton invariant mass $Q$,
using $y=2$ and $y=0$ for (a) and (b), respectively;
the dashed curve in (a) is the same as the solid curve in Figs.~\ref{fig:2}(a), (b).
For all cases in Fig.~\ref{fig:3}, we observe the typical flat behavior of $\aqt$ 
in the small $Q_T$ region, similarly as Fig.~\ref{fig:2}.
On the other hand, $\aqt$ increases for increasing $Q$, and the value
in the flat region reaches about 10\% for $Q=20$ GeV in Fig.~\ref{fig:3}(a).
Such dependence on $Q$ is associated with the small-$x$ behavior of the
relevant parton distributions:
smaller $Q$ corresponds to smaller $x_{1,2}^0 = e^{\pm y}Q/\sqrt{S}$, 
so that the small-$x$ rise of the unpolarized sea-distributions 
enhances the denominator of (\ref{asym}).
We obtain larger $\aqt$ for $y=2$ compared with the $y=0$
case, but the $y$-dependence of $\aqt$ is not so strong for all $Q$. 
For comparison, we also evaluate the $Q_T$-independent asymmetry
$A_{TT}$ of (\ref{eq:att})
including the NLO QCD corrections and with the same nonperturbative inputs 
as those used in Fig.~\ref{fig:3}. The results are shown in Table \ref{tab:1},
and these exhibit similar behavior with respect to the $Q$ and $y$ dependence
as that in Fig.~\ref{fig:3}. Note that we reproduce the NLO value of
$A_{TT}$ in Table \ref{tab:1} when we integrate respectively the
numerator and the denominator of the NLL+LO asymmetry (\ref{asym}) 
over $Q_T$ for each curve of Fig.~\ref{fig:3} 
(see discussion below (\ref{LOcross section})).
%
\begin{table}
\renewcommand{\arraystretch}{1.2}
\begin{tabular}{|c|c|c|c|c|c|c|}
\hline
\multicolumn{2}{|c|}{ } & $Q=2$GeV & $Q=5$GeV & $Q=8$GeV & $Q=15$GeV & $Q=20$GeV\\
\hline
$\sqrt{S}=200$GeV & $y=2$ & 3.3\% & 4.0\% & 4.9\% & 6.5\% & 7.4\% \\
\cline{2-7}
                  & $y=0$ & 3.5\% & 3.7\% & 4.4\% & 5.9\% & 6.9\% \\
\hline
$\sqrt{S}=500$GeV & $y=2$ & 1.8\% & 2.0\% & 2.4\% & 3.4\% & 4.0\% \\
\cline{2-7}
                  & $y=0$ & 2.2\% & 2.1\% & 2.4\% & 3.2\% & 3.8\% \\
\hline
\end{tabular}
\caption{The $Q_T$-independent asymmetry $A_{TT}$ of (\ref{eq:att})
including the NLO QCD corrections at RHIC kinematics.}
\label{tab:1}
\end{table}
But the NLO $A_{TT}$ 
are smaller by about 20\% than the corresponding values of the NLL+LO $\aqt$ 
in the ``flat'' region at small $Q_T$.
This enhancement of $\aqt$ compared with $A_{TT}$ arises from the soft
gluon resummation at the NLL level, as discussed in Fig.~\ref{fig:2}(a) above.

\begin{figure}
\vspace{0.5cm}
\bc
\includegraphics[height=5.8cm]{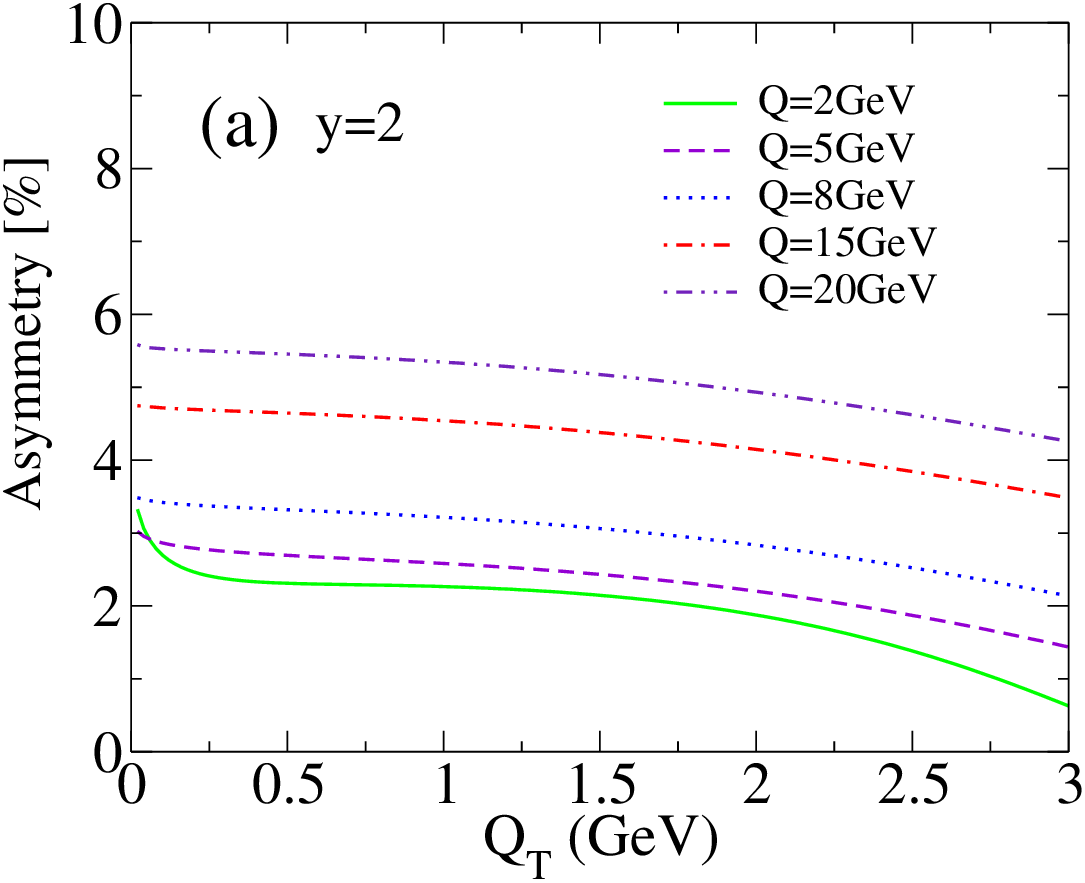}~~~~~~~
\includegraphics[height=5.8cm]{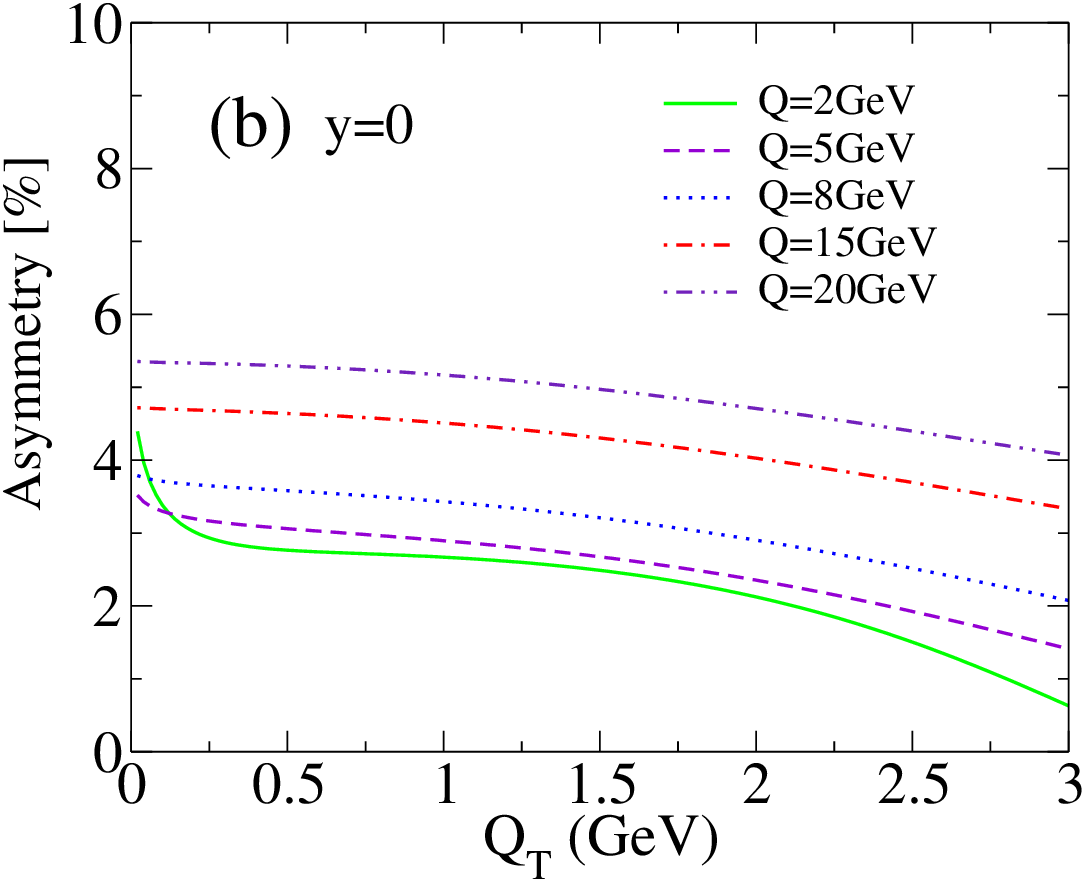}
\ec
\caption{Same as Fig.~\ref{fig:3}, but for $\sqrt{S}=500$ GeV.
}
\label{fig:4}
\end{figure} 

Fig.~\ref{fig:4} is same as Fig.~3, but for another RHIC kinematics with 
$\sqrt{S}=500$ GeV. 
General behavior for the $Q_T$, $Q$ and $y$ dependence is similar as
that in Fig.~\ref{fig:3}.
Comparing the curves with the same values of $Q$, $y$ between
Figs.~\ref{fig:3} and \ref{fig:4},
$\aqt$ are smaller for higher energy $\sqrt{S}=500$ GeV than those for 
$\sqrt{S}=200$ GeV.
This reflects the smaller $x_{1,2}^0 = e^{\pm y}Q/\sqrt{S}$ for larger $\sqrt{S}$, 
and the corresponding enhancement 
of the denominator in (\ref{asym}).
Similarly as Fig.~\ref{fig:3}, the NLL+LO $\aqt$ in the flat region of Fig.~\ref{fig:4}
are larger by 20-30\% than the corresponding NLO $A_{TT}$ shown in Table \ref{tab:1}.
It is generally true, 
regardless of the specific kinematics or the detailed behavior of
nonperturbative inputs, 
that the NLL+LO $\aqt$ of (\ref{asym}) in the flat region is considerably larger
than the corresponding NLO $A_{TT}$,
because this phenomenon is mainly governed by the partonic mechanism associated 
with the soft gluon resummation 
at the NLL level, as demonstrated in Figs.~\ref{fig:1}, \ref{fig:2}.
On the other hand, apparently the absolute magnitude of both $\aqt$ and $A_{TT}$
is influenced by the detailed behavior of the input parton distributions,
in particular, by their small-$x$ behavior at RHIC.
For example, if we change the input parton distributions, explained above (\ref{eq:np}),
from the NLO GRV98 and GRSV2000 distributions into
the NLO GRV94 \cite{GRV:94} and GRSV96 \cite{GRSV:96} distributions, 
the NLO values of $A_{TT}$ become smaller by 30-40\% 
than the corresponding values in Table \ref{tab:1}.
We note that the latter distributions are the ones used in the
calculation of \cite{MSSV:98}, 
and the small-$x$ behavior 
of the transversity distributions,
resulting from $\delta q(x,\mu^2_0)=[q(x,\mu^2_0)+\Delta
q(x,\mu^2_0)]/2$ at the input scale $\mu_0$, is 
rather different between those two choices of the distributions, 
reflecting that the helicity distributions at small $x$ 
are still poorly determined from experiments.~\footnote{
We thank H.~Yokoya and W.~Vogelsang for clarifying this point.}

\begin{figure}
\bc
\includegraphics[height=5.8cm]{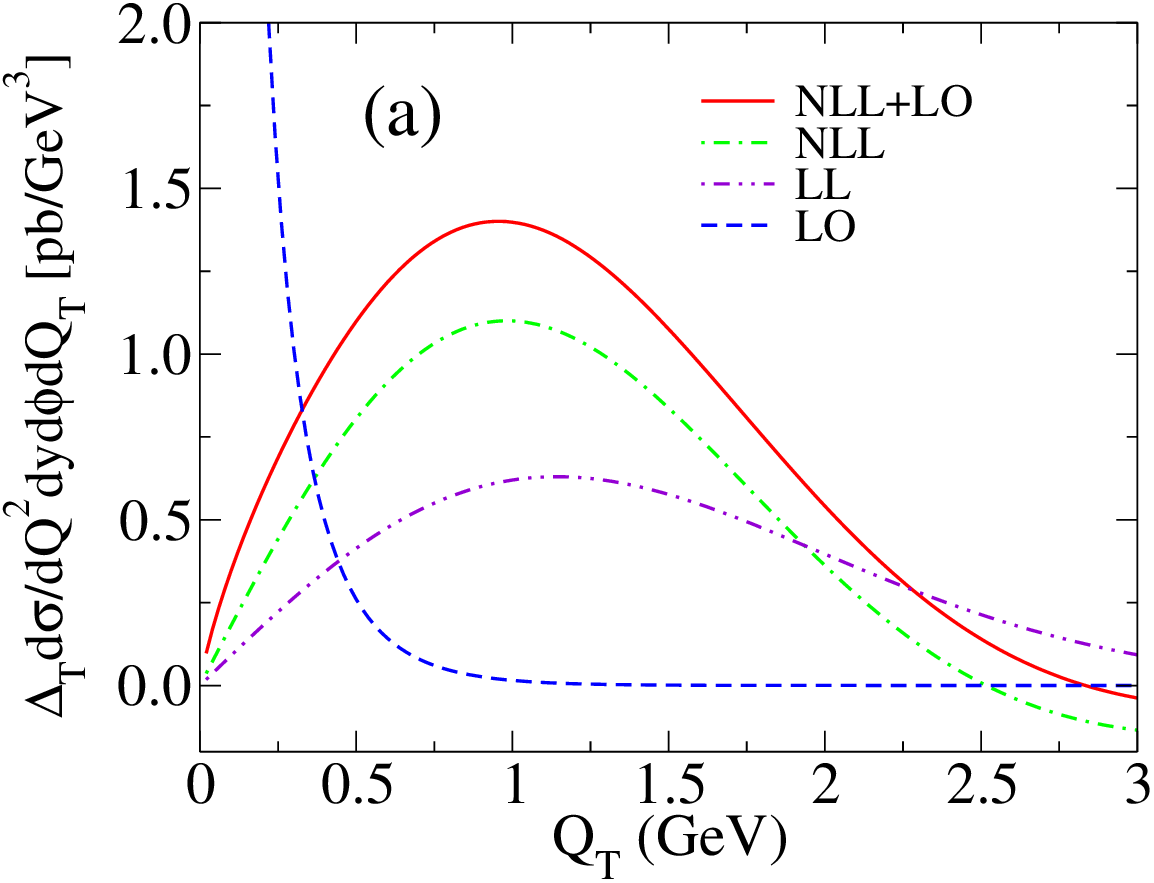}~~~~
\includegraphics[height=5.8cm]{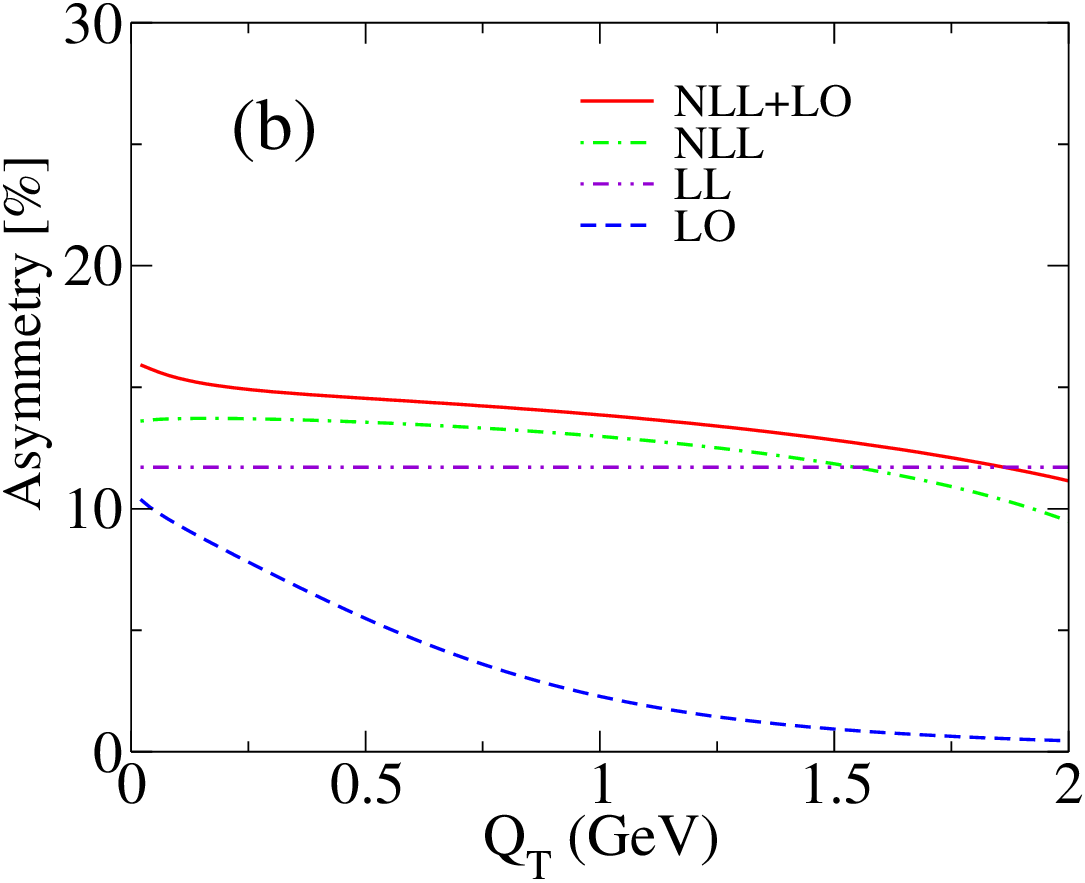}
\ec
\caption{The tDY at J-PARC kinematics, $\sqrt{S}=10$ GeV, $Q=2$ GeV, $y=0$ and $\phi=0$.
(a) The spin-dependent differential 
cross section $\Delta_Td\sigma/dQ^2 dQ_T dy d\phi$ using $g_{NP}=0.5$ GeV$^2$. 
(b) The asymmetries $\aqt$ obtained by using each curve in (a).
}
\label{fig:5}
\end{figure}

Next we discuss tDY foreseen at J-PARC.
Fig.~\ref{fig:5}(a) shows the $Q_T$ spectrum of the produced lepton pair 
for J-PARC kinematics,
$\sqrt{S}=10$ GeV, $Q=2$ GeV and $y =0$.
The curves show the spin-dependent differential cross sections, and 
have the same meaning as the corresponding curves in Fig.~\ref{fig:1}(a).
The double transverse-spin asymmetries are obtained as the ratio of the results 
in Fig.~\ref{fig:5}(a) to the corresponding results for the unpolarized 
differential cross sections, as shown in Fig.~\ref{fig:5}(b). 
We see that the results at J-PARC obey the similar pattern as those at RHIC shown 
in Figs.~\ref{fig:1}, \ref{fig:2}:
the flat behavior is observed for the NLL+LO $\aqt$
at $Q_T \rightarrow 0$ as well as around the peak region 
of the NLL+LO cross section,
and this is dominated by the NLL resummed components.
Also the soft-gluon resummation contributions 
at the NLL level enhances 
the asymmetry at the small $Q_T$-region significantly, 
compared with the LL of (\ref{LL}) 
or the fixed-order LO result.
As a result, we get $\aqt\simeq 15$\% as the NLL+LO prediction around the flat region,
which should be compared with the corresponding prediction 
$A_{TT}= 12.8$\% for (\ref{eq:att})
including the NLO corrections (see Table \ref{tab:2}). 
The reason why we obtain much larger values of $\aqt$, and also of
$A_{TT}$, than the RHIC case
is the larger $x_{1,2}^{0}=0.2$ probed at J-PARC, where the transversities are larger
and the unpolarized sea distributions are smaller.
Another difference compared with the RHIC case is that the contribution of
the ``regular component'' $\Delta_T \tilde{Y}$ of (\ref{matching}) 
in Fig.~\ref{fig:5}(a),
and the associated increase of the solid curve as $Q_T \rightarrow 0$ 
in Fig.~\ref{fig:5}(b),
due to the terms $\propto \ln (Q^2 /Q_T^2)$ in $\Delta_T \tilde{Y}$ and
$\tilde{Y}$ of (\ref{asym}), are more pronounced, but the latter effect shows up
only for $Q_T \lesssim 0.5$ GeV.

\begin{figure}
\bc
\includegraphics[height=5.8cm]{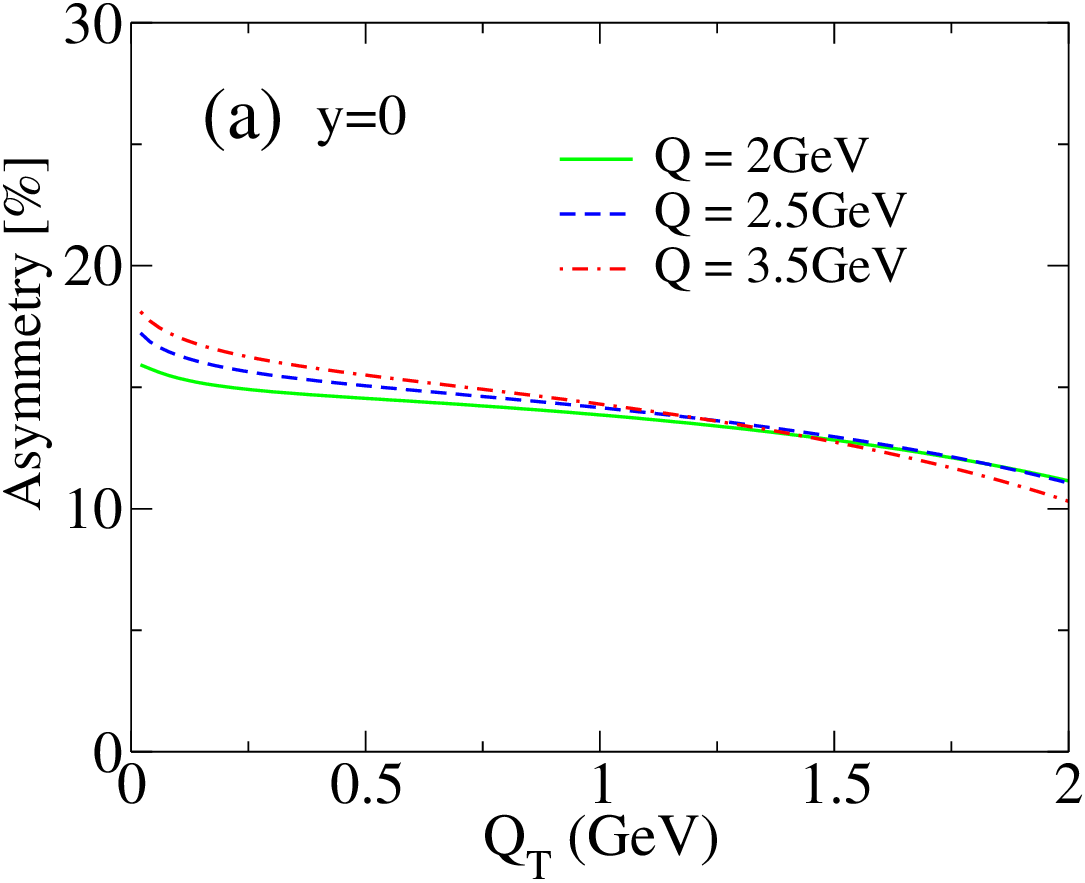}~~~~~~~~
\includegraphics[height=5.8cm]{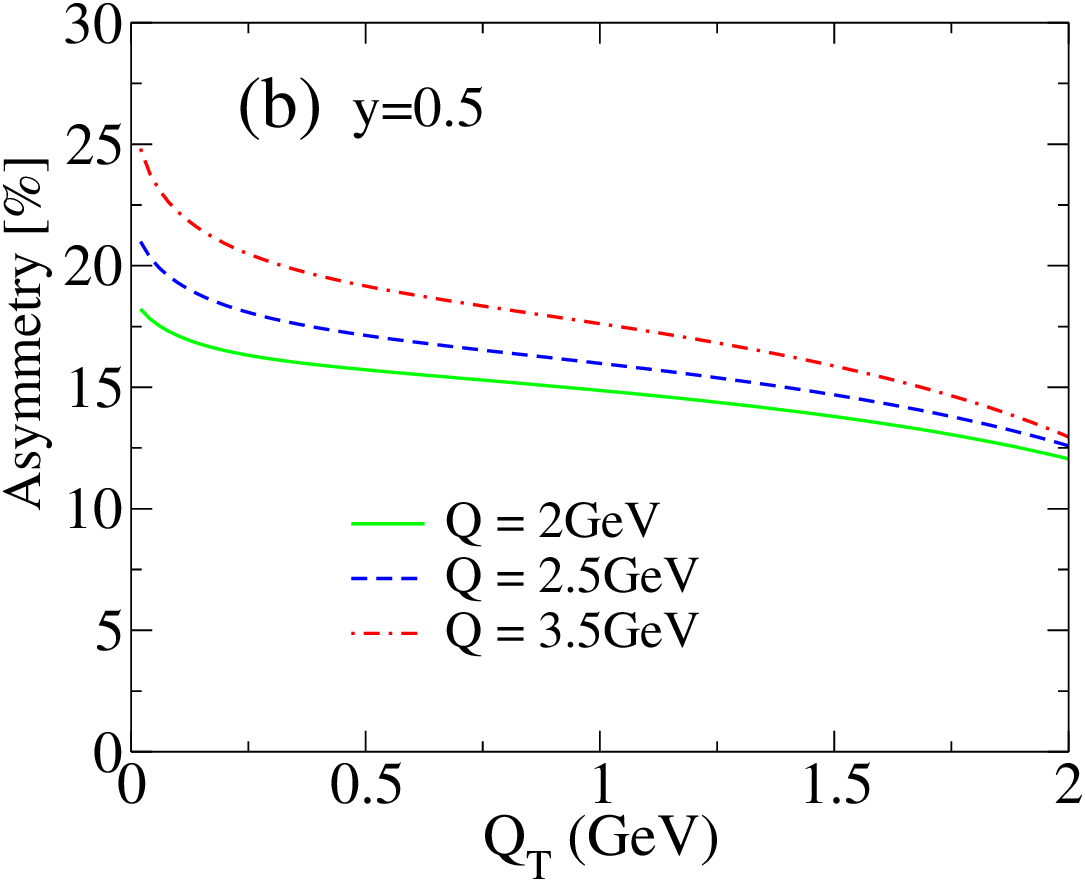}
\ec
\caption{The NLL+LO $\aqt$ of (\ref{asym}) with (\ref{eq:np}) using $g_{NP}=0.5$ GeV$^2$
at J-PARC kinematics, $\sqrt{S}=10$ GeV, $\phi=0$ with $y=0$ and $y=0.5$ for (a) and (b),
respectively.}
\label{fig:6}
\end{figure}

\begin{table}
\begin{center}
\renewcommand{\arraystretch}{1.2}
\begin{tabular}{|c|c|c|c|c|}
\hline
\multicolumn{2}{|c|}{ } & $Q=2$GeV & $Q=2.5$GeV & $Q=3.5$GeV \\
\hline
$\sqrt{S}=10$GeV & $y=0$   & 12.8\% & 12.9\% & 12.5\% \\
\cline{2-5}
                 & $y=0.5$ & 13.9\% & 14.8\% & 15.9\% \\
\hline
\end{tabular}
\end{center}
\caption{Same as Table \ref{tab:1} but for J-PARC kinematics.}
\label{tab:2}
\end{table}

In Fig.~\ref{fig:6} we show the NLL+LO asymmetries $\aqt$ of
(\ref{asym}) at J-PARC kinematics, $\sqrt{S}=10$ GeV and various values of $Q$,
with $y=0$ and $y=0.5$ for (a) and (b), respectively;
the solid curve in (a) is the same as the solid curve in Fig.~\ref{fig:5}(b).
We observe the flat behavior of $\aqt$ in the small $Q_T$ region, 
where $\aqt \simeq 15$-20\% and these values are significantly larger than
the corresponding results for the $Q_T$-independent, NLO asymmetry
$A_{TT}$ of (\ref{eq:att}), shown in Table \ref{tab:2}.
We note that the dependence of $\aqt$, as well as $A_{TT}$, on $Q$ is
weak in contrast to the RHIC case;
recall that the rather strong $Q$-dependence in Figs.~\ref{fig:3}, \ref{fig:4} 
was induced mainly by the growth of the unpolarized sea-distributions 
for the small $x_{1,2}^0$, probed at RHIC.

\section{The saddle point formula}

In Sec.~3, we have observed 
the universal flat behavior of NLL+LO $\aqt$ of (\ref{asym}) at small $Q_T$,
including the region around the peak of each DY cross section in the numerator and
denominator of (\ref{asym}) for both RHIC and J-PARC cases.
We have also demonstrated in Figs.~\ref{fig:2} and \ref{fig:5} 
that those flat behavior is driven by the dominant effects from soft gluon resummation 
embodied by the NLL resummed components 
$\Delta_T \tilde{X}^{\rm NLL}$ and $\tilde{X}^{\rm NLL}$
in (\ref{asym}).
As a result, the values of  
$\aqt$ obtained in the ``flat region'' of the corresponding 
experimental data may be
compared, to a good accuracy, with (\ref{asymNLL}).
Still, extraction of transversity distributions through such analysis 
should be a complicated task compared with the usual fixed-order analysis:
in the flat region of $\aqt$, 
the $b$ integration in the resummed part (\ref{resum}) 
(see also (\ref{resum:2}), (\ref{resum:21}))
mixes up the parton distributions numerically with 
very large perturbative effects due to the Sudakov factor shown 
in Figs.~\ref{fig:1} and \ref{fig:5},
as well as with another nonperturbative effects associated with $g_{NP}$ 
of (\ref{eq:np}).
Thus in each of the numerator and the denominator of (\ref{asym}),
the information on the parton distributions is associated with a portion 
of the large numerical quantity whose major part would cancel 
in the ratio of (\ref{asym}), and this fact would obscure 
the straightforward extraction of the transversity using
the above formulae like (\ref{resum:2}), (\ref{resum:21}), in particular with
respect to its accuracy.

We are able to derive a simple analytic formula which allows more direct extraction of
the transversity distributions 
from the experimental data in the flat region of $\aqt$ and also clarifies 
the accuracy of the resulting distributions.
For this purpose, we first note that the extrapolation of $\aqt$
in the flat region to $Q_T = 0$ corresponds to the case without 
the (experimentally uninteresting) weak enhancement at very small $Q_T$ due to the 
terms $\propto \log(Q^2/Q_T^2)$ 
in the regular components $\Delta_T \tilde{Y}$ and $\tilde{Y}$ of (\ref{asym}), 
so that the resulting value is very close
to the $Q_T \rightarrow 0$ limit of (\ref{asymNLL}) in
both RHIC and J-PARC cases (see Figs.~\ref{fig:2}-\ref{fig:6}).
Namely, $\aqtn^{\rm NLL}(Q_T=0)$ may be considered
to give a practical estimate of the data of $\aqt$ in the flat region
with a good accuracy. 
Then, at $Q_T = 0$, the region $|b| \sim 1/ \Lambda_{\rm QCD}$ becomes important for 
the $b$ integration of the relevant resummation formula (\ref{resum:21}).
Note that we can treat ``safely'' 
such long-distance region, corresponding to the boundary 
of perturbative and nonperturbative physics,
owing to that
the nonperturbative smearing (\ref{eq:np}) suppresses the too long-distance region
$|b| \gg 1/\Lambda_{\rm QCD}$, and that the dependence on a specific choice of $g_{NP}$
cancels in the asymmetries $\aqtn^{\rm NLL}(Q_T=0)$ as demonstrated in 
Fig.~\ref{fig:2}(b) (see also (\ref{eq:attnll}) below).

For simplicity in the presentation, we fix as $\mu_R=Q$ in the following.
In the relevant region $|b| \sim 1/ \Lambda_{\rm QCD}$,
we have 
$|\tilde{L}| \sim \ln(Q^2/\Lambda_{\rm QCD}^2 ) \sim 1/\alpha_s(Q^2 )$, i.e.,
$|\lambda| \sim 1$ (see (\ref{replaceL}), (\ref{eq:lambda})).
Because all logarithms, $\tilde{L}$ and $\ln (Q^2 /\Lambda^2_{\rm QCD})$, 
are counted equally large for $Q \gg \Lambda_{\rm QCD}$,
the resulting contributions to (\ref{resum:21}) 
are organized in terms of a single small parameter, $\alpha_s(Q^2 )$,
but with a different classification of the contributions in the order of $\alpha_s(Q^2 )$
from the usual perturbation theory that can be used in an other region, 
$0 \le |b| \lesssim 1/Q$:
as discussed below (\ref{eq:lambda}), when $\lambda = {\cal O}(1)$, 
the NLL contributions in the Sudakov exponent (\ref{sudakov:1})
produce the ${\cal O}(1)$ effects in the resummation formula (\ref{resum:21}),
while the NNLL contributions
could yield the corrections of ${\cal O}(\alpha_s(Q^2))$.
Therefore, when the region $|b| \sim 1/ \Lambda_{\rm QCD}$ is relevant 
as $Q_T \rightarrow 0$ 
and we neglect the NNLL contributions in (\ref{resum:21}),
the other contributions that correspond to the same order of 
$\alpha_s(Q^2)$ in (\ref{resum:2}) should be neglected
for a consistent treatment,
as $[1+ \alpha_s(Q^2) C_F (\pi^2 -8 )/2\pi] \rightarrow 1$, so that~\footnote{
In principle, we should use this classification also for the numerical
calculations presented in Sec.~3 when $Q_T \approx 0$. But 
we did not make the corresponding
replecement for the coefficient functions $C_{ij}$ at $Q_T \approx 0$ 
in the calculations of Figs.~\ref{fig:1}-\ref{fig:6}. If we performed that replacement,
the NLL+LO (\ref{asym}) as well as NLL (\ref{asymNLL}) 
asymmetries at $Q_T \approx 0$ in those figures would increase by about 5\%.}
\begin{equation}
\Delta_T X^{\rm NLL}_{N_1, N_2}  (Q_T^2 =0 , Q^2)
=
\delta H_{N_1,N_2}(Q^2)
I_{N_1,N_2}(Q_T^2=0, Q^2)\ .
\label{resum:3} 
\end{equation}
We note that the contributions to 
the NLL resummation formula for the unpolarized case can be classified 
similarly in the  $Q_T \rightarrow 0$ limit;
in particular, the present classification implies that
the gluon distributions decouple for $Q_T \rightarrow 0$ 
by negelecting the ${\cal O}(\alpha_s (Q^2))$ contributions 
of the corresponding 
coefficient functions $C_{ij}$ (see (\ref{UNPOL}) in Appendix).
 
It is also worth noting that this classification coincides with 
the ``degree 0 approximation'' discussed in \cite{CSS:85}: in general,
if one wants to evaluate the cross section for $Q_T \approx 0$
in an approximation where any corrections 
are suppressed by a factor of $[\ln (Q^2 /\Lambda^2_{\rm QCD})]^{-(N+1)}$,
one needs a ``degree $N$'' approximation; i.e., for the perturbatively 
calculable functions
in the general form of resummation formula (\ref{resum}) with (\ref{sudakov}), 
one needs $A_q$ to order $\alpha_s^{N+2}$,
$B_q$ to order $\alpha_s^{N+1}$, $C_{ij}$ to order $\alpha_s^{N}$, 
and the $\beta$ function to order $\alpha_s^{N+2}$.
This indicates that the NLL accuracy for a resummation formula
corresponds to the degree 0 approximation when the region $Q_T \approx 0$
is considered. In particular, this implies that 
the ${\cal O}(\alpha_s)$ contribution in the coefficient 
function $C_{ij}$ should be neglected for $Q_T \approx 0$; on the other hand, 
that contribution is necessary
to ensure the NLL accuracy for $Q_T \gtrsim \Lambda_{\rm QCD}$
in the classification based on resummed perturbation theory of 
towers of logarithms, 
$\alpha_s^n \ln^{2n-1}(Q^2/Q_T^2 )/Q_T^2$, 
$\alpha_s^n \ln^{2n-2}(Q^2/Q_T^2 )/Q_T^2$, 
and $\alpha_s^n \ln^{2n-3}(Q^2/Q_T^2 )/Q_T^2$ \cite{CSS:85,BCDeG:03,KKST:06}. 

Now we evaluate the $b$ integral of (\ref{resum:21})
at $Q_T =0$ according to the above classification.
We have to use the exponentiated form in the integrand of (\ref{resum:21})
without Taylor expansion,
because the region $|\lambda| \sim 1$ is relevant 
(see (\ref{sudakov:1})-(\ref{LO-evol})) \cite{KS:99}.
This type of integrals 
can be evaluated with the saddle point method: we extend the
saddle point evaluation applied to the LL resummation formula with 
$g_{NP} =0$ \cite{PP,CSS:85} into the case of our NLL resummation 
formula~(\ref{resum:21}) with nonzero $g_{NP}$. 
The corresponding extension is possible based on
the present formalism that accomplishes resummation at the partonic level.
We note that previous saddle-point calculations 
consider the case with $g_{NP}=0$ to avoid model dependence 
for the prediction of the cross sections,
but the resultant saddle-point formula is applicable only to the production
of extremely high-mass DY pair
and is practically useless (see e.g. \cite{PP,CSS:85,QZ}).
We find that, with nonzero $g_{NP}$, we can obtain a new saddle-point formula
applicable to the RHIC and J-PARC cases; also, although the behavior of 
the cross sections are influenced 
by specific value of $g_{NP}$, the asymmetries are not,
as already noted above.

When $Q$ is large enough, $\alpha_s (Q^2) \ll 1$, so that the $b$
integral in (\ref{resum:21}) with (\ref{eq:np}) and $Q_T =0$ is dominated 
by a saddle point determined by mainly the LL
term in the exponent (\ref{sudakov:1}) of the Sudakov form factor~\cite{PP,CSS:85}.
In this case, the contributions to the $b$ integration from too short 
($|b| \ll 1/Q$) and long distance ($|b| \gg 1/\Lambda_{\rm QCD}$)
along the integration contour ${\cal C}$ explained below (\ref{replaceL}) are
exponentially suppressed: 
this allows us to give up the replacement (\ref{replaceL});
also we may neglect the integration along the two branches, 
$b=b_c + e^{\pm i\theta}t$ with $t \in \{0, \infty \}$, in  ${\cal C}$, when 
$b_c$ is sufficiently large but is 
less than the position of the singularity in the Sudakov exponent, $b_L$.
In fact, we can check numerically that the relevant 
integrand has a nice saddle point well below $b_L$ (above $0$) for the kinematics 
of our interest.
Then, changing the integration variable to 
$\lambda$, given by (\ref{eq:lambda}),
we get (see (\ref{eq:h0}), (\ref{sudakov:2}), (\ref{LO-evol}))
\begin{equation}
I_{N_1 , N_2}(Q_T^2 = 0, Q^2)=
\frac{b_0^2}{4Q^2 \beta_0 \alpha_s(Q^2)}\int_{-\infty}^{\lambda_c}  d\lambda 
e^{-\zeta(\lambda)+ h^{(1)}(\lambda) + R_{N_1}(\lambda) 
+ R_{N_2}(\lambda)}\ ,
\label{eq:sp}
\end{equation}
where $\lambda_c = \beta_0 \alpha_{s}(Q^2) \ln(Q^2 b_c^2/b_0^2 )$ ($<1$), and 
\begin{equation}
\zeta(\lambda) = - \frac{\lambda}{\beta_0 \alpha_s(Q^2)}
-\frac{h^{(0)}(\lambda)}{\alpha_s (Q^2)}
+ \frac{g_{NP}b_0^2}{Q^2}e^{\frac{\lambda}{\beta_0 \alpha_s (Q^2 )}}\ .
\label{eq:fxi}
\end{equation}
An important point is that the ratio,
$[ h^{(1)}(\lambda) + R_{N_1}(\lambda) + R_{N_2}
(\lambda)]/\zeta(\lambda)$, actually behaves as a quantity of the order 
of $\alpha_s(Q^2 )$
in the relevant region $0<\lambda < \lambda_c$ of the integration 
in (\ref{eq:sp}), even for nonzero $g_{NP} \simeq 0.5$ GeV$^2$.
The precise position of the saddle point in the integral of (\ref{eq:sp}) 
is determined by the condition,
$- {\zeta}'(\lambda)+ {h^{(1)}}'(\lambda ) + {R_{N_1}}'(\lambda) 
+ {R_{N_2}}' (\lambda) = 0$, and we express its solution as
$\lambda = \lambda_{SP} + \Delta \lambda_{SP}$ where $\lambda_{SP}$ is the 
solution of $\zeta'(\lambda)=0$, i.e., 
\begin{equation}
1-\frac{A_q^{(1)}}{2\pi \beta_0}\frac{\lambda_{SP}}{1-\lambda_{SP}} = 
\frac{g_{NP}b_0^2}{Q^2}e^{\frac{\lambda_{SP}}{\beta_0 \alpha_s (Q^2 )}}
\label{eq:lsp}
\end{equation}
is satisfied, and 
$\Delta \lambda_{SP}= 
[{h^{(1)}}' (\lambda_{SP}) + {R_{N_1}}' (\lambda_{SP}) 
+ {R_{N_2}}' (\lambda_{SP}) ]
/{\zeta}''(\lambda_{SP})$ 
denotes the shift of the saddle point at the NLL accuracy.
Evaluating (\ref{eq:sp}) around 
$\lambda= \lambda_{SP} + \Delta \lambda_{SP}$, 
we get 
\begin{equation}
I_{N_1 , N_2}(0, Q^2)=\left(
\frac{b_0^2}{4Q^2 \beta_0 \alpha_s(Q^2)} \sqrt{\frac{2\pi}{\zeta''(\lambda_{SP})}} 
e^{-\zeta(\lambda_{SP})+h^{(1)}(\lambda_{SP})}
\right)e^{R_{N_1}(\lambda_{SP}) + R_{N_2}(\lambda_{SP})},
\label{eq:speval}
\end{equation}
to the NLL accuracy. Here the contributions from the third or higher order terms
in the Taylor expansion of the exponent in (\ref{eq:sp}) about the saddle point 
$\lambda= \lambda_{SP} + \Delta \lambda_{SP}$, 
as well as the other terms generated by the shift $\Delta \lambda_{SP}$, are
found to give the effects behaving as ${\cal O} (\alpha_s(Q^2 ) )$, i.e,
are of the same order as the NNLL corrections, and thus are neglected, 
similarly as in (\ref{resum:2}), (\ref{resum:3}),
according to the classification of the contributions at $Q_T=0$. 
Substituting (\ref{eq:speval}) into (\ref{resum:3})
and performing the double inverse Mellin transformation to the ($x_1^0$, $x_2^0$) space,
the result is expressed as the factor in the parentheses of (\ref{eq:speval}),
multiplied by (\ref{tPDF}) with the scale,
$\mu_F \rightarrow b_0/b_{SP}$ where $b_{SP}= (b_0 /Q)e^{\lambda_{SP}/(2\beta_0
\alpha_s(Q^2))}$, 
because $e^{R_{N_1}(\lambda_{SP})}$, $e^{R_{N_2}(\lambda_{SP})}$ 
in (\ref{eq:speval}) 
can be identified with the NLO evolution operators from the scale $Q$ 
to $b_0/b_{SP}$, to the present accuracy (see (\ref{LO-evol})).
The saddle-point evaluation of the corresponding resummation formula for 
the unpolarized case can be performed similarly, and the result 
is given by the above result for the polarized case, with the replacement 
$\delta H ( x_1^0, x_2^0;\ b_0^2 /b_{SP}^2 ) \rightarrow 
H ( x_1^0, x_2^0;\ b_0^2 /b_{SP}^2 )$.
The common factor for both the polarized and unpolarized results, 
given by the contribution in the parentheses of (\ref{eq:speval}),
involves ``very large perturbative effects'' due to the Sudakov factor,
and shows the well-known asymptotic behavior \cite{PP},
$\sim (\Lambda_{\rm QCD}^2 /Q^2 )^{a\ln(1+1/a)}$ 
with $a\equiv A_q^{(1)}/(2\pi \beta_0)$,
for $Q \gg \Lambda_{\rm QCD}$; 
but this factor cancels out for the asymmetry.
As a result, we obtain the $Q_T\rightarrow 0$ limit of (\ref{asymNLL}) as
\begin{equation}
\aqtn^{\rm NLL}(Q_T =0)=\frac{1}{2}\cos(2 \phi ) 
\frac{\delta H \left( x_1^0, x_2^0;\ b_0^2 /b_{SP}^2 \right)}
{H \left( x_1^0, x_2^0;\ b_0^2 /b_{SP}^2 \right)}\ ,
\label{eq:attnll}
\end{equation}
which is exact, up to the NNLL corrections corresponding to 
the ${\cal O}(\alpha_s (Q^2) )$ effects.
This remarkably compact formula is reminiscent of 
$\aqtn^{\rm LL}(Q_T)$ of (\ref{LL}) that retains only the LL level resummation,
or the $Q_T$ independent asymmetry of (\ref{eq:att}),
but is different in the scale of the parton distributions from those 
leading-order results.
Namely, our result (\ref{eq:attnll}) demonstrates:
in the $Q_T =0$ limit,
the all-order soft-gluon-resummation effects on the asymmetry 
mostly cancel between the numerator and the denominator 
of (\ref{eq:attnll}), but certain contributions at the NLL level survive 
the cancellation and 
are entirely absorbed into the unconventional scale $b_0/b_{SP}$ for the
relevant distribution functions.

The new scale $b_0/b_{SP}$ is determined by solving (\ref{eq:lsp})
numerically, substituting
$A_q^{(1)}=2C_F$ from (\ref{eq:AB}) and input values for $Q$ and $g_{NP}$,
but it is useful to consider its general behavior: the LHS of (\ref{eq:lsp}) equals 1 at 
$\lambda_{SP}=0$, decreases as a concave function for increasing $\lambda_{SP}$,
and vanishes at $\lambda_{SP}=1/[1+A_q^{(1)}/(2 \pi \beta_0 )] \cong 0.6$; while
the RHS is in general much smaller than 1 at $\lambda_{SP}=0$, 
increases as a convex function for increasing $\lambda_{SP}$,
and is larger than 1 at $\lambda_{SP} \simeq 1$.
Thus the solution of (\ref{eq:lsp}) corresponds to the case with 
${\rm LHS}={\rm RHS}\simeq 1/2$,
more or less independently of the specific value of $Q$ and $g_{NP}$, so that we get
$b_0 /b_{SP} \simeq b_0 \sqrt{2g_{NP}}$. 
This result depends only mildly on the nonperturbative parameter $g_{NP}$,
and suggests that one may always use $b_0 /b_{SP} \simeq 1$ GeV,
for the cases of our interest where $Q$ is of several GeV and $g_{NP} \simeq 0.5$ GeV$^2$
as in Figs.~\ref{fig:1}-\ref{fig:6}.
The actual numerical solution of (\ref{eq:lsp}) justifies this simple consideration at 
the level of 20\% accuracy.
This fact will be particularly helpful in the first attempt to compare 
(\ref{eq:attnll}) with the experimental data so as to
extract the transversity distributions.

Our saddle-point formula (\ref{eq:attnll}) embodies the characteristic features 
of the NLL soft gluon resummation effects on the asymmetries $\aqt$, 
emphasized in Sec.~3.
In particular, our derivation of (\ref{eq:attnll}) demonstrates clearly the mechanism,
which makes the parton distributions at the low scale $\sim Q_T$ 
play dominant roles,
and leads to the ``enhancement'' of the dot-dashed curve 
in Figs.~\ref{fig:2} and \ref{fig:5}.
As noted in the beginning of this section, (\ref{eq:attnll}) may be
directly compared with the experimental value 
of the asymmetries $\aqt$, observed around the peak of the $Q_T$ spectrum of 
the corresponding DY cross sections. 
But there is one caution for such application.
As seen from the above derivation, the parton distributions appearing
in (\ref{eq:attnll}) are the NLO distributions up to the corrections 
at the NNLL level; e.g., the transversity distributions appearing in the
numerator of (\ref{eq:attnll}) is obtained by evolving the customary NLO 
transversity $\delta q(x, Q^2)$ at the scale $Q$,
to the scale $b_0 /b_{SP}$ using (\ref{LO-evol}) that is 
{\it the NLO evolution operators up to the NNLL corrections}.
Therefore, the formula (\ref{eq:attnll}) can be used
in the region where NNLL corrections are small;
we know that the NNLL corrections at $Q_T \approx 0$ 
correspond to ${\cal O}(\alpha_s (Q^2 ))$ effects, 
and should be negligible in general. 
However, such straightforward estimate might fail 
at the edge region of the phase space, e.g., at the small $x$ region:
because the relevant evolution operators (\ref{LO-evol}) actually coincide with 
the leading contributions in the large-logarithmic expansion of the usual
LO DGLAP evolution,\footnote{This fact also suggests that one may use the fixed value,
$b_0 /b_{SP} \simeq 1$ GeV, in (\ref{eq:attnll}) for all $Q$ (and $g_{NP}$)
rather than solving (\ref{eq:lsp}) numerically for each different 
input value of $Q$, $g_{NP}$, 
because the sensitivity of the LO evolution on the small change of the scale is modest.
} 
(\ref{eq:attnll}) would not accurate when the NLO corrections
in the usual DGLAP evolution are large compared with 
the contributions of (\ref{LO-evol}).
Such situation would typically occur in the region with small $x_{1,2}^0$,
corresponding to the case with large $\sqrt{S}$.
In Table~\ref{tab:3}, we compare $\aqtn^{\rm NLL}(Q_T =0)$
using the numerical $b$-integration (``NB''),
obtained as the $Q_T \rightarrow 0$ limit of the dot-dashed curve
in Figs~\ref{fig:3}(a) and \ref{fig:6}(a),
with those using the saddle-point formula (\ref{eq:attnll}).
For the latter we use $b_0 /b_{SP}$ obtained as the
solution of (\ref{eq:lsp}) with $g_{NP}=0.5$ GeV$^2$, and consider the two cases 
for the parton distributions participating in (\ref{eq:attnll}): 
``SP-I'' uses the parton distributions 
which are obtained by evolving the customary NLO distributions at the scale $Q$,
to $b_0 /b_{SP}$ using the NLO evolution operators up to
the NNLL corrections like (\ref{LO-evol});
``SP-II'' uses the customary NLO distributions at the scale $b_0 /b_{SP}$.
Here the ``customary NLO distributions'' are constructed 
as described above (\ref{eq:np}).
First of all, the results for SP-I
demonstrate the remarkable accuracy of our simple analytic formula 
(\ref{eq:attnll}) for both RHIC and J-PARC, reproducing the results of NB 
to the 10\% accuracy.~\footnote{If we use the fixed value,
$b_0 /b_{SP} = 1$ GeV, for all cases, instead of the solution of (\ref{eq:lsp}), 
the results in SP-I change by at most 5\%, for both RHIC and J-PARC kinematics.
The corresponding change in SP-II is by less than 5\% for J-PARC,
and by about 10\% (15\%) for $Q =2$-8 GeV ($Q=15$-20 GeV) at RHIC.
}
On the other hand, the results for SP-II indicate that the NNLL
corrections are moderate for large $\sqrt{S}$ at RHIC, while those are
expected to be small for small $\sqrt{S}$ at J-PARC.

We propose that our simple formula (\ref{eq:attnll}) is 
applicable to the analysis of low-energy experiment at J-PARC in
order to extract the NLO transversity distributions directly from the data.  
On the other hand, (\ref{eq:attnll}) will not be so accurate for analyzing 
the data at RHIC, but will be still useful for
obtaining the first estimate of the transversities. 
We emphasize that such (moderate) uncertainty in applying our formula
(\ref{eq:attnll}) to the RHIC case is not caused by the saddle-point evaluation, 
nor by considering the $Q_T \rightarrow 0$ limit, but rather
is inherent in the general $Q_T$ resummation framework
which, at the NLL level, implies the use of the evolution operators
(\ref{LO-evol}) with the LO DGLAP kernel;
more accurate treatment of the small-$x$ region of the parton distributions 
relevant to the RHIC case would require the resummation formula to the
NNLL accuracy, where the NLO DGLAP kernel participates
in the evolution operators (\ref{LO-evol}) 
from $Q$ to $b_0 /b$ (see e.g. \cite{BCDeG:03}).

\begin{table}
\renewcommand{\arraystretch}{1.2}
\begin{tabular}{|c|r|r|r|r|r||r|r|r|}
\hline
& \multicolumn{5}{|c||}{$\sqrt{S}=200$ GeV, \hspace{0.1cm} $y=2$}&
\multicolumn{3}{|c|}{$\sqrt{S}=10$ GeV,  \hspace{0.1cm} $y=0$}\\
\hline
 $Q$  &  2GeV  &  5GeV  & 8GeV & 15GeV & 20GeV 
& 2GeV & 2.5GeV  & 3.5GeV \\
\hline
SP-I
& 4.3\% & 5.4\% & 6.6\% & 8.7\% & 9.8\% 
& 14.1\% & 14.5\% & 14.8\%  \\
SP-II
& 7.3\% & 8.7\% & 9.8\% & 11.8\% & 12.7\%  
& 14.7\% & 14.8\% & 14.2\% \\
NB
& 3.8\% & 4.9\%  &  6.1\% & 8.2\% & 9.4\% 
& 13.4\% & 14.0\% & 14.9\% \\
\hline
\end{tabular}
\caption{The $Q_T \rightarrow 0$ limit of $\aqtn^{\rm NLL}(Q_T)$ of (\ref{asymNLL})
for RHIC and J-PARC kinematics. SP-I and SP-II are the results of the 
saddle-point formula (\ref{eq:attnll}) for $g_{NP}=0.5$ GeV$^2$,
using the evolution operators from $Q$ to $b_0 /b_{SP}$, to the NLL accuracy
 and to the customary NLO accuracy, respectively. NB is obtained from 
the dot-dashed curve in Figs~\ref{fig:3}(a) and \ref{fig:6}(a).}
\label{tab:3}
\end{table}

\section{Conclusions}
In this paper we have presented a study of double transverse-spin
asymmetries for dilepton production at small $Q_T$ in $pp$ collisions. 
The logarithmically enhanced contributions, which arise in the
small $Q_T$ region due to multiple soft gluon emission in QCD,
are resummed to all orders in $\alpha_s$ 
up to the NLL accuracy.
Based on this framework, we calculate numerically the spin-dependent and 
spin-averaged cross sections in tDY, and the corresponding asymmetries $\aqt$,
as a function of $Q_T$ at RHIC kinematics as well as at J-PARC kinematics.
The soft gluon resummation contributions make the cross sections finite 
and well-behaved over all
regions of $Q_T$, so that the singular $Q_T$ spectra in the fixed-order
perturbation theory are redistributed,
forming a well-developed peak in the small $Q_T$ region.
As a result, 
both the polarized and unpolarized cross sections become more ``observable'' 
around the pronounced ``peak region'' at small $Q_T$, involving the bulk of events.
Reflecting the universal nature of the soft gluon effects,
those large resummation-contributions mostly 
cancel in the cross section asymmetries $\aqt$, leading to the almost
constant behavior of $\aqt$ in the small $Q_T$ region,
but, remarkably, the effects surviving the cancellation
raise the corresponding constant value of $\aqt$ considerably compared with
the asymmetries in the fixed-order
perturbation theory.
We have obtained a QCD prediction as $\aqt \simeq 5$-10\% and 15-20\% 
in the ``flat region'' 
for typical kinematics at RHIC and J-PARC, respectively, where
the different values of $\aqt$ are associated with 
the different values of parton's momentum fraction probed by these two experiments. 

We have also derived a new saddle-point formula for $\aqtn(Q_T \approx 0)$,
clarifying the classification of the contributions 
involved in the resummation formula for 
$Q_T \rightarrow 0$.
The formula is exact to the NLL accuracy, and 
embodies the above remarkable features of soft gluon resummation 
effects at small $Q_T$ 
in a compact analytic form.
Our saddle-point formula may be compared with the data of $\aqt$ 
in the peak region of the DY $Q_T$-spectrum, and thus 
provides us with a new direct approach to extract the transversity distributions from
experimental data.

We mention that there is 
another kind of logarithmically enhanced soft-gluon contributions, 
subject to the so-called
``threshold resummation'',  besides those treated by the $Q_T$ resummation. 
It is known that the threshold resummation effects on the cross sections 
can be important when the probed momentum fractions of partons are
rather large like at J-PARC.
The corresponding effects for tDY are studied 
\cite{SSVY:05} in $p\bar{p}$ collisions at GSI kinematics, and 
the results indicate that the threshold resummation effects will not be so
significant for the kinematical regions corresponding to experiments 
at J-PARC, and, furthermore, will cancel mostly in the asymmetries.

We have revealed that
the ``amplification'' of the double transverse-spin asymmetries $\aqt$ at small $Q_T$
is driven by the partonic mechanism participating at the NLL level,
as the interplay between 
the large logarithmic gluon effects resummded into the universal Sudakov factor
and the DGLAP evolutions specific for each channel.
Thus similar phenomenon is anticipated also 
in $p\bar{p}$ collisions at the future experiments at
GSI \cite{PAX:05}, where the large values are predicted 
for the $Q_T$ independent asymmetry $A_{TT}$ \cite{SSVY:05,BCCGR:06}.
The application of our $Q_T$ resummation formalism 
to $p\bar{p}$ collision will be presented elsewhere \cite{KKT:07}. 

\section*{Acknowledgments}
We thank Werner Vogelsang, Hiroshi Yokoya and Stefano Catani 
for useful discussions and comments. 
The work of J.K. and K.T. was 
supported by the Grant-in-Aid 
for Scientific Research Nos. C-16540255 and C-16540266. 

\appendix
\section*{Appendix: Resummed cross section for unpolarized DY}
In this appendix, we summarize the corresponding results
for the unpolarized Drell-Yan process which are necessary
to calculate the asymmetry $\aqt$.
Although all of them have already appeared in the 
literatures~\cite{AEGM:84,CSS:85},
we will list them for the convenience of the reader.

The NLL resummed component, corresponding to (\ref{resum}) for the polarized case, 
reads,
\bea
\lefteqn{X^{\rm NLL} (Q_T^2, Q^2, y)
   = 
        \int_0^{\infty} d b \, \frac{b}{2}\,
     J_0 (b q_T)\, e^{\, S (b , Q)}\  
      \biggl[ H (x_1^0\,,\,x_2^0\,;\, b^2_0 / b^2 )}  \nonumber\\
    &+& \frac{\alpha_s(b^2_0/b^2)}{2 \pi}
     \left\{ 
          \int_{x_1^0}^1 \frac{dz}{z} \, C_{qq}^{(1)} (z)\,
                    H (x_1^0 / z \,,\,x_2^0\,;\, b^2_0 / b^2 ) 
          + \int_{x_2^0}^1 \frac{dz}{z} \, C_{qq}^{(1)} (z)\,
                    H (x_1^0 \,,\,x_2^0 / z \,;\, b^2_0 / b^2 ) \right. \nonumber\\
    &+& \left.
          \int_{x_1^0}^1 \frac{dz}{z} \, C_{qg}^{(1)} (z)\,
                    K_2 (x_1^0 / z \,,\,x_2^0\,;\, b^2_0 / b^2 ) 
          + \int_{x_2^0}^1 \frac{dz}{z} \, C_{qg}^{(1)} (z)\,
                    K_1 (x_1^0 \,,\,x_2^0 / z \,;\, b^2_0 / b^2 ) 
                   \right\}\, \biggr]\ ,
\label{UNPOL}
\eea
where $S(b,Q)$ is the exponent of the Sudakov factor, given by 
(\ref{sudakov})-(\ref{eq:lambda}), 
and 
\bean
    C_{qq}^{(1)} (z) &=& C_{\bar{q}\bar{q}}^{(1)} (z) 
        = C_F \, (1 - z) + C_F \left( \frac{\pi^2}{2} - 4 \right)\,
                   \delta (1 - z) \ ,\\
    C_{qg}^{(1)} (z) &=& C_{\bar{q} g}^{(1)} (z)
        = 2\, T_R\, z (1-z) \ ,
\eean
with $T_R = 1/2$. The parton distributions are defined as,
\bean
   H (x_1\,,\,x_2;\, \mu_F^2) &=& \sum_q\, e_q^2\,
\left[ q (x_1, \mu_F^2)\, \bar{q} (x_2, \mu_F^2)
+\bar{q} (x_1, \mu_F^2)\, q (x_2, \mu_F^2)\right]\ ,\\
   K_1 (x_1\,,\,x_2;\, \mu_F^2) &=& \sum_q\, e_q^2\, 
              \left[ q (x_1, \mu_F^2) + \bar{q} (x_1, \mu_F^2) \right]
                         \, g (x_2, \mu_F^2) \ ,\\
   K_2 (x_1\,,\,x_2;\, \mu_F^2) &=& \sum_q\, e_q^2\, g(x_1, \mu_F^2)\,
              \left[ q(x_2, \mu_F^2) + \bar{q}(x_2, \mu_F^2) \right]\ .
\eean
The finite component, corresponding to $\Delta_T Y$ of 
(\ref{cross section}) for the polarized case,
is given by,
\bean
Y (Q_T^2\,,\,Q^2\,,\,y )=
        \frac{\alpha_s (\mu_R^2)}{2 \pi}\, \left[ C_F\, Y_q (Q_T^2\,,\,Q^2\,,\,y )
  + T_R\, Y_g (Q_T^2\,,\,Q^2\,,\,y ) \right] \ ,
\eean
where the first term comes from the $q\bar{q}$ annihilation subprocess as
\bean
  Y_q (Q_T^2\,,\,Q^2\,,\,y )
     &=& -\, \frac{2}{S} \left[ 
            \int^1_{\sqrt{\tau_+} \, e^y} \, \frac{d x_1}{x_1 - x_1^+} 
         \ \frac{H (x_1\,,\,x_2^* ;\, \mu_F^2)}{x_1 x_2^*}
     + \int^1_{\sqrt{\tau_+} \, e^{-y}} \, \frac{d x_2}{x_2 - x_2^+} 
         \ \frac{H (x_1^*\,,\,x_2 ;\, \mu_F^2)}{x_1^* x_2} \right]\\
     &+& \frac{1}{Q_T^2}\,
      \left[ \int d x_1 \, H_1 + \int d x_2 \, H_2 
                        + 2\, H (x_1^0\,,\,x_2^0 ;\, \mu_F^2)
      \, \ln \frac{(1 - x_1^+)(1 - x_2^+)}{(1 - x_1^0)(1 - x_2^0)} 
           \right] \ ,  
\eean
with the variables according to
\cite{AEGM:84,KKST:06}:\footnote{Note that $x_{1\,,\,2}^+ = \sqrt{\tau_+} e^{\pm y}=
x_{1\,,\,2}^* = x_{1\,,\,2}^0$ when $Q_T^2 = 0$.}
\bean
    x_1^+ 
       &=& \left( \frac{Q^2 + Q^2_T}{S}\right)^{1/2} e^y \ ,\quad
    x_2^+ 
       = \left( \frac{Q^2 + Q^2_T}{S}\right)^{1/2} e^{- y}\ ,\\
x_1^* &=& \frac{x_2 x_1^+ - x_1^0 x_2^0}{x_2 - x_2^+}\ , \quad
    x_2^* = \frac{x_1 x_2^+ - x_1^0 x_2^0}{x_1 - x_1^+}\ , \quad
    \sqrt{\tau_+} = \sqrt{\frac{Q_T^2}{S}} +
                    \sqrt{\tau + \frac{Q_T^2}{S}}\ .
\eean
We used the shorthand notation for the integral that vanishes for $Q_T=0$,
\bean
   \int d x_1 \, H_1 &\equiv& \int^1_{\sqrt{\tau_+} \, e^y}\ 
      \,  \frac{d x_1}{x_1 - x_1^+} 
        \left[ H (x_1\,,\,x_2^* ;\, \mu_F^2)
       \left\{ 1 + \left(\frac{\tau}{x_1 x_2^*}\right)^2 \right\}
            - 2\, H (x_1^0 \,,\,x_2^0 ;\, \mu_F^2) \right]\\
   && \qquad\qquad
    - \int^1_{x_1^0}\ 
      \,  \frac{d x_1}{x_1 - x_1^0} 
        \left[ H (x_1\,,\,x_2^0 ;\, \mu_F^2)
       \left\{ 1 + \left (\frac{x_1^0}{x_1} \right)^2 \right\}
        - 2\, H (x_1^0 \,,\,x_2^0 ;\, \mu_F^2) \right]\ ,  
\eean
and $\int d x_2 \, H_2 = \left. \int d x_1 \, H_1 \right|_{y \rightarrow -y}$.
For the contribution from the gluon Compton process, we have
\bean
  Y_g (Q_T^2\,,\,Q^2\,,\,y )
     &=& \frac{1}{S} \left[
       \int^1_{\sqrt{\tau_+} \, e^y} \, \frac{d x_1}{x_1 - x_1^+} 
    \ K_1 (x_1\,,\,x_2^* ;\, \mu_F^2) \frac{x_1 x_2^+ - \tau}{(x_1 x_2^*)^2} \right.\\
    &&  \qquad\qquad\qquad\qquad + \,
     \left. \int^1_{\sqrt{\tau_+} \, e^{-y}} \, \frac{d x_2}{x_2 - x_2^+} 
         \ K_1 (x_1^*\,,\,x_2 ;\, \mu_F^2) \frac{x_1^* x_2^+ - \tau}{(x_1^* x_2)^2}
                            \right]\\
     &+& \frac{1}{Q_T^2}\,
      \left[ \int^1_{\sqrt{\tau_+} \, e^{-y}} \, \frac{d x_2}{x_2 - x_2^+} 
         \ K_1 (x_1^*\,,\,x_2 ;\, \mu_F^2) \left\{ 
           \frac{x_2 x_1^+ - \tau}{x_1^* x_2} -
                 \frac{2\, \tau \, (x_2 x_1^+ - \tau)^2}
                 {(x_1^* x_2)^3} \right\} \right.\\
     && \qquad\qquad\qquad\qquad - 
       \, \int^1_{x_2^0} \, \frac{d x_2}{x_2} 
         \ K_1 (x_1^0\,,\,x_2 ;\, \mu_F^2) \left\{ 
           1  -  \frac{2\, x_2^0 \, (x_2 - x_2^0)}
                 {x_2^2} \right\} \\
     && \quad + \left.
         \int^1_{\sqrt{\tau_+} \, e^{y}} \, \frac{d x_1}{x_1 - x_1^+} 
         \ K_1 (x_1\,,\,x_2^* ;\, \mu_F^2) \left\{ 
           \frac{x_2^* x_1^+ - \tau}{x_1 x_2^*} -
                 \frac{2\, \tau \, (x_2^* x_1^+ - \tau)^2}
                 {(x_1 x_2^*)^3} \right\} \right] \\
     &+& ( 1 \leftrightarrow 2) \ ,
\eean
where $1 \leftrightarrow 2$ means the exchange of the suffix of variables
as well as $y \leftrightarrow -\, y$.
As for the parton distributions, this exchange should be read,
\[  K_1 (x_1\,,\,x_2^*) \leftrightarrow K_2 (x_1^*\,,\,x_2)\ ,
   \quad  K_1 (x_1^*\,,\,x_2) \leftrightarrow K_2 (x_1\,,\,x_2^*)\ ,
   \quad  K_1 (x_1^0\,,\,x_2) \leftrightarrow K_2 (x_1\,,\,x_2^0)\ .\]
The fixed-order cross section in the $\overline{\rm MS}$ scheme is given as
\begin{eqnarray}
\frac{d \sigma^{\rm FO}}{d Q^2 d Q_T^2 d y d \phi}
= 
\frac{2\alpha^2}{3\, N_c\, S\, Q^2}
 \left[ X\, (Q_T^2 \,,\, Q^2 \,,\, y) 
+ Y\, (Q_T^2 \,,\, Q^2 \,,\, y) \right],
\label{cross sectionU}
\end{eqnarray}
to the ${\cal O}(\alpha_s)$ accuracy, where 
$X\, (Q_T^2 \,,\, Q^2 \,,\, y) \equiv \left. X^{\rm NLL}\, (Q_T^2 \,,\, Q^2 \,,\, y) \right|_{\rm FO}$
denotes the ``singular'' part
resulting from the expansion of the resummed component
up to the fixed-order $\alpha_s(\mu_R^2 )$. 
The formula (\ref{cross sectionU}) should be compared with
(\ref{cross section}).
Taking the similar steps as those in (\ref{eq:MT})-(\ref{matching}),
we obtain the differential cross section 
for unpolarized DY with the soft gluon resummation as  
\bea
\frac{d\sigma}{dQ^2dQ_T^2dyd\phi}=
\frac{2\alpha^2}{3\, N_c\, S\, Q^2}
\biggl[ \tilde{X}^{\rm NLL}(Q_T^2 , Q^2,y)
+ \tilde{Y}(Q_T^2 , Q^2,y)\biggr],
\label{NLL+LOU}
\eea
which is used to calculate (\ref{asym}).


\begin{thebibliography}{99}
\bibitem{BDR:02}
See e.g., 
V.~Barone, A.~Drago and P.~G.~Ratcliffe, Phys.~Rep.
{\bf 359}, (2002) 1;
J.~Kodaira and K.~Tanaka,
Prog.\ Theor.\ Phys.\  {\bf 101} (1999) 191.

\bibitem{Anselmino:07}
M.~Anselmino, et al., Phys. Rev. {\bf D75} (2007) 054032.

\bibitem{Ralston:1979ys}
J.~P.~Ralston and D.~E.~Soper,
Nucl.\ Phys.\  {\bf B152} (1979) 109;
  R.~L.~Jaffe and X.~D.~Ji,
Nucl.\ Phys.\  {\bf B375} (1992) 527.

\bibitem{MSSV:98}
O.~Martin, A.~Sch\"afer, M.~Stratmann and W.~Vogelsang, 
Phys. Rev. {\bf D57} (1998) 3084; 
ibid. {\bf D60} (1999) 117502.

\bibitem{WV:98}
A.~Mukherjee, M.~Stratmann and W.~Vogelsang,
Phys.\ Rev.\  {\bf D67} (2003) 114006;
A.~Mukherjee, M.~Stratmann and W.~Vogelsang,
Phys.\ Rev.\  {\bf D72} (2005) 034011.

\bibitem{KKST:06}
H.~Kawamura, J.~Kodaira, H.~Shimizu and K.~Tanaka,
Prog. Theor. Phys. {\bf 115} (2006) 667.

\bibitem{Dutta}
D.~Dutta et al., Letter of Intent (L15) for Nuclear and Particle Physics 
Experiments at J-PARC, 
http://www-ps.kek.jp/jhf-np/LOIlist/LOIlist.html.

\bibitem{KNV:06}
Y.~Koike, J.~Nagashima and W.~Vogelsang,
Nucl.\ Phys.\  {\bf B744} (2006) 59.

\bibitem{CDL:06}
M.~Contalbrigo, A.~Drago and P.~Lenisa, hep-ph/0607143, in Proceedings
of XIV International Workshop on Deep Inelastic Scattering (DIS2006),
Tsukuba, Japan, eds. M.~Kuze et al. (World Scientific, 2007), p.727. 

\bibitem{DS:84}
C.~T.~H.~Davies and W.~J.~Stirling, Nucl.~Phys. {\bf B244} (1984) 337. 

\bibitem{KT:82}
J.~Kodaira and L.~Trentadue, Phys. Lett. {\bf B112} 
(1982) 66 ; Report {\bf SLAC-Pub-2934} (1982); 
Phys. Lett. {\bf B123} (1983) 335.

\bibitem{dG}
D. de Florian and M. Grazzini,
Phys. Rev. Lett. {\bf 85} (2000) 4678.

\bibitem{LKSV:01} 
E.~Laenen, G.~Sterman and W.~Vogelsang, Phys. Rev. 
{\bf D63} (2001) 114018;
S.~Kulesza, G.~Sterman and W.~Vogelsang, Phys. Rev. 
{\bf D66} (2002) 014011. 

\bibitem{BCDeG:03}
S.~Catani, D.~de Florian and M.~Grazzini,
  Nucl.\ Phys.\  {\bf B596} (2001) 299;
G.~Bozzi, S.~Catani, D.~de Florian and M.~Grazzini, 
Phys. Lett. {\bf B564} (2003) 65; Nucl. Phys. {\bf B737} 
(2006) 73.

\bibitem{CSS:85} 
J.~C.~Collins, D.~Soper and  G.~Sterman, 
Nucl. Phys. {\bf B250} (1985) 199.

\bibitem{AM:90}
X.~Artru and M.~Mekhfi, Z.~Phys. {\bf C45} (1990) 669.

\bibitem{KMHKKV:97}
A.~Hayashigaki, Y.~Kanazawa and Y.~Koike, 
Phys. Rev. {\bf D56} (1997) 7350;
S.~Kumano and M.~Miyama, Phys. Rev. {\bf D56} (1997) 2504;
W.~Vogelsang, Phys. Rev. {\bf D57} (1998) 1886.

\bibitem{AEGM:84}
G.~Altarelli, R.~K.~Ellis, M.~Greco and G.~Martinelli,  
Nucl. Phys. {\bf B246} (1984) 12.

\bibitem{Soffer:95}
J.~Soffer, Phys.~Rev.~Lett. {\bf 74} (1995) 1292.

\bibitem{GRV:98}
M.~Gl\"uck, E.~Reya and A.~Vogt, Eur.~Phys.~J. 
{\bf C5} (1998) 461. 

\bibitem{GRSV:00}
M.~Gl\"uck, E.~Reya, M.~Stratmann, and W.~Vogelsang, 
Phys.~Rev. {\bf D63} (2001) 094005.

\bibitem{KS:03} 
A.~Kulesza and W.~J.~Stirling, JHEP {\bf 0312} (2003) 056.

\bibitem{GRV:94}
M.~Gluck, E.~Reya and A.~Vogt,
Z.\ Phys.\  {\bf C67} (1995) 433.

\bibitem{GRSV:96}
M.~Gluck, E.~Reya, M.~Stratmann and W.~Vogelsang,
Phys.\ Rev.\  {\bf D53} (1996) 4775.

\bibitem{PP}
G.~Parisi and R.~Petronzio,
Nucl.\ Phys.\  {\bf B154} (1979) 427;
J.~C.~Collins and D.~E.~Soper,
Nucl.\ Phys.\  {\bf B197} (1982) 446.

\bibitem{KS:99} 
A.~Kulesza and W.~J.~Stirling,
Nucl.\ Phys.\  {\bf B555} (1999) 279.

\bibitem{QZ}
J.~Qiu and X.~Zhang,
Phys.\ Rev.\ Lett.\  {\bf 86} (2001) 2724.

\bibitem{SSVY:05}
H.~Shimizu, G.~Sterman, W.~Vogelsang and H.~Yokoya, 
Phys.~Rev. {\bf D71} (2005) 114007.

\bibitem{PAX:05}
V.~Barone, et al.[PAX Collabolation], hep-ex/0505054;
M.~Maggiora et al. [ASSIA Collabolation], hep-ex/0504011. 

\bibitem{BCCGR:06}
V.~Barone, A.~Caferella, C~.Coriano, M.~Guzzi and P.~G.~Ratcliffe, 
Phys.~Lett {\bf B639} (2006) 483.

\bibitem{KKT:07}
H.~Kawamura, J.~Kodaira and K.~Tanaka, in preparation.

\end{thebibliography}
\end{document}